\documentclass[acmsmall,screen]{acmart}

\usepackage{microtype}

\usepackage{listings}
\definecolor{strings}{rgb}{.624,.251,.259}
\definecolor{keywords}{rgb}{.224,.451,.686}
\definecolor{comment}{rgb}{.5,.5,.5}
\lstdefinelanguage{loopnest}
  {morekeywords={allocate,compute,for,@blocks,
                 @shared_mem,@threads,@serial,
                 unrolled,in,Func,Var},
   sensitive=true,
   morecomment=[l]{--},
   morecomment=[s]{--[[}{]]},
   morestring=[b]",
   morestring=[d]',
   keywordstyle=\color{keywords},
   stringstyle=\color{strings},
   commentstyle=\color{comment}
}

\usepackage{letltxmacro}
\newcommand*{\SavedLstInline}{}
\LetLtxMacro\SavedLstInline\lstinline
\DeclareRobustCommand*{\lstinline}{%
  \ifmmode
    \let\SavedBGroup\bgroup
    \def\bgroup{%
      \let\bgroup\SavedBGroup
      \hbox\bgroup
    }%
  \fi
  \SavedLstInline
}
\newcommand{\code}{\lstinline}
\newcommand{\secref}[1]{Sec.~\ref{#1}}
\newcommand{\figref}[1]{Fig.~\ref{#1}}

\newcommand{\X}[0]{$\times$}
\newcommand{\Xs}[0]{$\times$ } % Add a space after X

\newcommand{\oneshotsamplingfreezingspeedup}[0]{1.22}
\newcommand{\topfivesamplingfreezingspeedup}[0]{1.27}
\newcommand{\meanstatesevaluated}[0]{5.19\%}
\newcommand{\beststatesevaluated}[0]{0.69\%}
\newcommand{\geomeancompiletimealloff}[0]{370} 
\newcommand{\geomeancompiletimeallon}[0]{7.6} 
\newcommand{\geomeancompiletimespeedup}[0]{49} 
\newcommand{\bestcompiletimespeedup}[0]{530}

\newcommand{\autotunespeedup}[0]{1.7} 
\newcommand{\humanspeedup}[0]{0.95} 

\newcommand{\compiletimessioutas}[0]{
\begin{table*}
    \centering
    
    \caption{Compile time comparison between our One Shot, Top 5, and Autotuned modes against Sioutas et al. \cite{Sioutas2020}. The times for One Shot and Top 5 are the median of 20 independent trials. For One Shot mode and Sioutas et al., autoscheduling times (i.e. compile time excluding code generation time) are given in parentheses.}
    
    \begin{tabular}{|l||c|c|c|c|}
\hline
\multicolumn{5}{|c|}{Compile Time (s) -- Ours vs. Sioutas et al. \cite{Sioutas2020} -- Lower is Better} \\
\hline
& Sioutas et al. & One Shot  & Top 5 & Autotuning \\
\hline

bilateral grid & 52 (43.8) & \textbf{23.5 (9.6)} & 34 & 677 \\
\hline
local laplacian & \textbf{61 (47.8)} & 149.7 (121.8) & 149.7 & 3627 \\
\hline
nl means & 57 (49.3) & \textbf{55.7 (41.9)} & \textbf{55.7} & 1104 \\
\hline
lens blur & \textbf{85 (71)} & 186.2 (149.1) & 186.2 & 4248 \\
\hline
camera pipe & \textbf{59} (58) & 69 (\textbf{57.5}) & 70 & 1609 \\
\hline
stencil chain & \textbf{100} (86.2) & 111.1 (\textbf{80.1}) & 113.5 & 2431 \\
\hline
harris & \textbf{6 (4.3)} & 22.1 (12.6) & 34 & 715 \\
\hline
histogram equalize & \textbf{6} (5.3) & 16.6 (\textbf{3.3}) & 32 & 667 \\
\hline
max filter & 32 (22.7) & \textbf{19.3 (7.6)} & 33 & 685 \\
\hline
unsharp mask & \textbf{4 (3.1)} & 16 (3.2) & 33 & 680 \\
\hline
interpolate & \textbf{28 (18.8)} & 50.1 (36.8) & 51 & 1180 \\
\hline
conv layer & \textbf{3 (1.2)} & 16.9 (4.8) & 31 & 620 \\
\hline
matrix multiply & \textbf{7} (0.7) & 12.6 (\textbf{0.1}) & 29 & 561 \\
\hline
IIR blur & \textbf{5} (4) & 15.7 (\textbf{2.7}) & 33 & 682 \\
\hline
BGU & \textbf{78 (56.7)} & 165.4 (131.4) & 165.4 & 4046 \\
\hline
depthwise sep. conv & \textbf{10 (2.3)} & 23.5 (12.7) & 41 & 808 \\
\hline
learned demosaic & \textbf{8 (1.8)} & 98.5 (82.2) & 98.5 & 2006 \\
\hline

\textbf{geomean} & \textbf{18.6 (11.3)} & 40.3 (15.7) & 55.8 & 1186.7\\
\hline

    \end{tabular}
    
    \label{fig:results-compile-times-sioutas}
\end{table*}
}

\newcommand{\compiletimestable}[0]{
\begin{table*}
    \centering
        \caption{Beam search compile times for all configurations of memoization, hierarchical sampling, and freezing on and off. We highlight the lowest number in each row in bold. Our optimizations, when used in conjunction, significantly improve compile times. In general, they have more impact on pipelines that favor schedules with many compute\_root stages, and less impact on pipelines that favor fusion, for example, camera pipe.}
    \begin{tabular}{|l||c|c|c|c|c|c|c|c|}
\hline
\multicolumn{9}{|c|}{Beam Search Compile Time (s) -- Lower is Better} \\
\hline
\textit{Memoization} & ON & ON  & ON & ON & OFF & OFF & OFF & OFF  \\
\textit{Sampling} & ON & ON & OFF & OFF & ON & ON & OFF & OFF  \\
\textit{Freezing} & ON & OFF & ON & OFF & ON & OFF & ON & OFF  \\
\hline
\hline

bilateral grid & \textbf{2.94} & 16.51 & 7.87 & 27.95 & 9.96 & 212.86 & 338.45 & 682.42 \\
\hline
local laplacian & \textbf{110.09} & 432 & 131.47 & 509.91 & 418.45 & 7918.44 & 3996.33 & 14568.95 \\
\hline
nl means & \textbf{11.16} & 23.04 & 26.61 & 34.85 & 45.77 & 201.11 & 269.01 & 536.12 \\
\hline
lens blur & \textbf{111.21} & 674.67 & 273.73 & 960.84 & 394.15 & 23379.77 & 15588.4 & 59085.78 \\
\hline
camera pipe & 42.91 & 719.92 & 463.62 & 1287.87 & \textbf{42.29} & 712.76 & 468.96 & 1280.13 \\
\hline
stencil chain & \textbf{52.9} & 183.35 & 74.1 & 216.81 & 225.45 & 2334.77 & 2238.13 & 2633.46 \\
\hline
harris & \textbf{4.52} & 13.73 & 8.67 & 20.3 & 32.92 & 14.94 & 149.88 & 57.73 \\
\hline
hist. equalize & \textbf{2.06} & 5.28 & 4.38 & 8.69 & 11.08 & 28.32 & 45.78 & 71.06 \\
\hline
max filter & \textbf{1.77} & 4.48 & 3 & 5.89 & 7.56 & 17.78 & 24.57 & 38.77 \\
\hline
unsharp mask & 2.56 & 8.04 & \textbf{2.12} & 10.26 & 5.47 & 8.88 & 17.52 & 17.99 \\
\hline
interpolate & \textbf{25.75} & 138.35 & 42.16 & 178.68 & 89.18 & 3170.35 & 1253.08 & 5211.48 \\
\hline
conv layer & \textbf{1.1} & 12.69 & 4.44 & 18.52 & 3.79 & 93.59 & 41.2 & 214.92 \\
\hline
matrix multiply & \textbf{0.08} & 0.24 & 0.22 & 0.48 & 0.1 & 0.27 & 1.36 & 0.57 \\
\hline
IIR blur & \textbf{0.6} & 1.51 & 1.08 & 2.05 & 2.43 & 8.18 & 8.51 & 20.26 \\
\hline
BGU & \textbf{89.39} & 246.48 & 146.15 & 272.64 & 388.56 & 2431.77 & 1836.45 & 3578.3 \\
\hline
dep. sep. conv & \textbf{4} & 10.52 & 11.72 & 15.74 & 8.43 & 73.49 & 113.41 & 175.68 \\
\hline
l. demosaic & \textbf{54.24} & 89.64 & 157.99 & 181.49 & 173.37 & 702.62 & 2589.62 & 2034.93 \\
\hline

\textbf{geomean} & \textbf{7.61} & 28.61 & 16.92 & 41.78 & 24.78 & 168.74 & 217.94 & 372.06 \\
\hline

    \end{tabular}

    \label{fig:results-compile-times}
\end{table*}
}

\newcommand{\compiletimespeeduptable}[0]{
\begin{table*}[h]
    \centering
        \caption{Beam search compile time speedup for all configurations of memoization, hierarchical sampling, and freezing on and off, over the configuration with all off. We highlight the highest number in each row in bold. With all on, there is a compile time speed up on average of \geomeancompiletimespeedup\Xs and in the best case of \bestcompiletimespeedup\X.}
    \begin{tabular}{|l||c|c|c|c|c|c|c|c|}
\hline
\multicolumn{9}{|c|}{Beam Search Compile Time Speedup -- Higher is Better} \\
\hline
\textit{Memoization} & ON & ON  & ON & ON & OFF & OFF & OFF & OFF  \\
\textit{Hierarchical Sampling} & ON & ON & OFF & OFF & ON & ON & OFF & OFF  \\
\textit{Freezing} & ON & OFF & ON & OFF & ON & OFF & ON & OFF  \\
\hline
\hline

bilateral grid & \textbf{232.32} & 41.33 & 86.66 & 24.41 & 68.53 & 3.21 & 2.02 & 1 \\
\hline
local laplacian & \textbf{132.33} & 33.72 & 110.82 & 28.57 & 34.82 & 1.84 & 3.65 & 1 \\
\hline
nl means & \textbf{48.03} & 23.26 & 20.15 & 15.39 & 11.71 & 2.67 & 1.99 & 1 \\
\hline
lens blur & \textbf{531.32} & 87.58 & 215.85 & 61.49 & 149.91 & 2.53 & 3.79 & 1 \\
\hline
camera pipe & \textbf{29.83} & 1.78 & 2.76 & 0.99 & 30.27 & 1.8 & 2.73 & 1 \\
\hline
stencil chain & \textbf{49.79} & 14.36 & 35.54 & 12.15 & 11.68 & 1.13 & 1.18 & 1 \\
\hline
harris & \textbf{12.77} & 4.2 & 6.66 & 2.84 & 1.75 & 3.86 & 0.39 & 1 \\
\hline
histogram equalize & \textbf{34.47} & 13.46 & 16.22 & 8.18 & 6.41 & 2.51 & 1.55 & 1 \\
\hline
max filter & \textbf{21.95} & 8.66 & 12.91 & 6.58 & 5.13 & 2.18 & 1.58 & 1 \\
\hline
unsharp mask & \textbf{7.04} & 2.24 & 8.47 & 1.75 & 3.29 & 2.03 & 1.03 & 1 \\
\hline
interpolate & \textbf{202.37} & 37.67 & 123.62 & 29.17 & 58.44 & 1.64 & 4.16 & 1 \\
\hline
conv layer & \textbf{195.76} & 16.94 & 48.39 & 11.6 & 56.67 & 2.3 & 5.22 & 1 \\
\hline
matrix multiply & \textbf{7.55} & 2.32 & 2.56 & 1.19 & 5.73 & 2.09 & 0.42 & 1 \\
\hline
IIR blur & \textbf{33.67} & 13.42 & 18.82 & 9.88 & 8.33 & 2.48 & 2.38 & 1 \\
\hline
BGU & \textbf{40.03} & 14.52 & 24.48 & 13.12 & 9.21 & 1.47 & 1.95 & 1 \\
\hline
depthwise separable conv & \textbf{43.95} & 16.7 & 14.99 & 11.16 & 20.83 & 2.39 & 1.55 & 1 \\
\hline
learned demosaic & \textbf{37.51} & 22.7 & 12.88 & 11.21 & 11.74 & 2.9 & 0.79 & 1 \\
\hline
\textbf{geomean} & \textbf{48.92} & 13.01 & 21.99 & 8.9 & 15.01 & 2.2 & 1.71 & 1 \\
\hline

    \end{tabular}

    \label{fig:results-compile-times-speedup}
\end{table*}
}
\newcommand{\memorytable}[0]{
\begin{table*}
    \caption{Memory usage with memoization both on and off, for a single beam search schedule, a single greedy schedule, and a full batch of 80 schedules. The numbers for One Shot and Top 5 are the median of 20 independent trials.}
    \centering
    \begin{tabular}{|l||c|c|c|c|c|c|}
\hline
\multicolumn{7}{|c|}{Memoization Memory Usage (MB)} \\
\hline
\textit{Memoization} & \multicolumn{3}{|c|}{ON} & \multicolumn{3}{|c|}{OFF}  \\
\hline
& Beam Search & Greedy & Batch & Beam Search & Greedy & Batch \\
\hline

bilateral grid & 113 & 110 & 8814 & 112 & 110  & 8799 \\
\hline
local laplacian & 419 & 126 & 10369 & 148 & 125 & 10023 \\
\hline
nl means & 1142 & 112 & 9969 & 269 & 110 & 8996 \\
\hline
lens blur & 691  & 165 & 13740 & 292 & 157 & 12719 \\
\hline
camera pipe & 118  & 112 & 8986 & 118 & 112 & 9001 \\
\hline
stencil chain & 144 & 136 & 10863 & 113 & 135 & 10760 \\
\hline
harris & 115  & 111 & 8862 & 115 & 111 & 8857  \\
\hline
histogram equalize & 113 & 111 & 8891 & 115 & 111 & 8882 \\
\hline
max filter & 154 & 111 & 8926 & 115 & 111 & 8887  \\
\hline
unsharp mask & 112 & 111 & 8854 & 112 & 111 & 8854  \\
\hline
interpolate & 166  & 113 & 9079 & 120 & 113 & 9034  \\
\hline
conv layer & 111 & 109 & 8722 & 112 & 109 & 8732 \\
\hline
matrix multiply & 108 & 103 & 8213 & 108 & 103 & 8213 \\
\hline
IIR blur & 111 & 109 & 8702 & 111 & 109 & 8706 \\
\hline
BGU & 18957 & 1590 & 144558 & 3861 & 1536  & 125210 \\
\hline
depthwise sep. conv & 520 & 110  & 9171 & 225  & 110 & 8891  \\
\hline
learned demosaic & 912 & 168 & 14214 & 275 & 151 & 12208 \\
\hline

    \end{tabular}

    \label{fig:results-memory-usage}
\end{table*}
}

\AtBeginDocument{%
  \providecommand\BibTeX{{%
    \normalfont B\kern-0.5em{\scshape i\kern-0.25em b}\kern-0.8em\TeX}}}

\acmSubmissionID{45}

\setcopyright{rightsretained}
\acmPrice{}
\acmDOI{10.1145/3485486}
\acmYear{2021}
\copyrightyear{2021}
\acmSubmissionID{oopsla21main-p45-p}
\acmJournal{PACMPL}
\acmVolume{5}
\acmNumber{OOPSLA}
\acmArticle{109}
\acmMonth{10}

\begin{document}


\title[Efficient Automatic Scheduling of Imaging and Vision Pipelines for the GPU]{Efficient Automatic Scheduling of Imaging and Vision Pipelines for the GPU}


\author{Luke Anderson}
\affiliation{%
  \institution{Massachusetts Institute of Technology}
  \country{USA}}
\email{lukea@mit.edu}

\author{Andrew Adams}
\affiliation{%
  \institution{Adobe}
  \country{USA}}

\author{Karima Ma}
\affiliation{%
  \institution{Massachusetts Institute of Technology}
  \country{USA}}

\author{Tzu-Mao Li}
\affiliation{%
  \institution{Massachusetts Institute of Technology \& University of California, San Diego}
  \country{USA}}

\author{Tian Jin}
\affiliation{%
  \institution{Massachusetts Institute of Technology}
  \country{USA}}

\author{Jonathan Ragan-Kelley}
\affiliation{%
  \institution{Massachusetts Institute of Technology}
  \country{USA}}

\renewcommand{\shortauthors}{L. Anderson, A. Adams, K. Ma, T.-M. Li, T. Jin, and J. Ragan-Kelley}

\begin{abstract}
We present a new algorithm to quickly generate high-performance GPU implementations of complex imaging and vision pipelines, directly from high-level Halide algorithm code. It is fully automatic, requiring no schedule templates or hand-optimized kernels.
We address the scalability challenge of extending search-based automatic scheduling to map large real-world programs to the deep hierarchies of memory and parallelism on GPU architectures in reasonable compile time.
We achieve this using (1) a two-phase search algorithm that first `freezes' decisions for the lowest cost sections of a program, allowing relatively more time to be spent on the important stages, (2) a hierarchical sampling strategy that groups schedules based on their structural similarity, then samples representatives to be evaluated, allowing us to explore a large space with few samples, and (3) memoization of repeated partial schedules, amortizing their cost over all their occurrences.
We guide the process with an efficient cost model combining machine learning, program analysis, and GPU architecture knowledge.

We evaluate our method's performance on a diverse suite of real-world imaging and vision pipelines. Our scalability optimizations lead to average compile time speedups of \geomeancompiletimespeedup\Xs (up to \bestcompiletimespeedup\X). We find schedules that are on average \autotunespeedup\Xs faster than existing automatic solutions (up to 5\X), and competitive with what the best human experts were able to achieve in an active effort to beat our automatic results.

\end{abstract}




\maketitle

\section{Introduction}  \label{sec:introduction}

There is an increasing demand for high-performance imaging and vision algorithms, but implementing these programs on GPUs involves making optimization choices from a large space of options (e.g., splitting and reordering loops and assigning them to GPU blocks and threads, fusing different stages into single GPU kernels, and caching intermediate results in shared memory).
Imaging and vision programs require particular attention to long-range fusion and trading redundant recomputation for reduced bandwidth and greater locality~\cite{halide2013}.
Manually exploring these options is time-consuming, and it is difficult to predict ahead of time whether a change will help or hurt performance. 

Our goal is to automatically optimize these programs. We build on Halide, a domain-specific compiler that decouples the algorithm -- \emph{what} to compute -- from the the schedule -- \emph{how} to compute it. This separation makes it easier to explore different schedules for a given algorithm, but finding high-performance schedules remains a challenge. The space of possible schedules for a given program is combinatorially large, so it is not feasible to compile and benchmark all of them. Instead, we seek a solution that can explore a large search space efficiently, and can evaluate the performance of potential options without needing to compile and benchmark every choice.

Previous approaches have focused largely on two disparate domains --- neural networks and imaging --- and developed specialized techniques for each.
Tensor compilers (e.g.,~\cite{Chen:2018:Learning,Vasilache:2018:TCF,Zheng:2020:FAS,Jia:2019:TOD,Zheng:2020:AGH}) focus on neural networks, which are dominated by dense, high arithmetic intensity kernels. These computations can be broken up into small components (such as a convolution layer or ResNet block) and the scheduling space within each operator explored independently. However, as we show in \secref{sec:graph-partitioning}, this technique is a poor fit for imaging pipelines, which must schedule tens or hundreds of stages jointly, applying long-range fusion for high performance.
At the same time, existing autoschedulers for imaging pipelines have achieved impressive results for some programs and target architectures, but face a scalability problem when applied directly to the larger scheduling space required to achieve peak performance on GPUs~\cite{Adams2019,Sioutas2020}.

We aim to efficiently explore a rich space of GPU schedules and achieve high performance on a broad range of imaging and vision applications. To achieve this in a scalable way, we develop three complementary strategies:

\begin{enumerate}

\item
We factor autoscheduling into a pre-pass, where the program is scheduled with a restricted set of search options and the cheapest stages of computation have their schedule options `frozen' accordingly. The remainder of the autoscheduling process spends relatively more time performing a more expansive search on just the important parts of the program.

\item 
We introduce a hierarchical sampling strategy which uses a structural hash to group similar schedules and evaluate only representatives of each group. This allows us to effectively explore a large space of possible schedules while only considering a small subset of options.

\item
We recognize and exploit repeated substructure within the space of schedules, memoizing the analysis of partial schedules to amortize the cost over many occurrences.
\end{enumerate}
These optimizations significantly reduce the number of states evaluated, leading to an average compile time speedup of \geomeancompiletimespeedup\X{} (up to \bestcompiletimespeedup\X). At the same time, they also better stratify the search space, yielding schedules that are on average $1.2\times$ faster than with the optimizations disabled.

Even with all three optimizations, there are still orders of magnitude too many choices to compile and benchmark each one. We extend the method of Adams et al.~\cite{Adams2019} and use a learned cost model to guide our search for the best schedule. To balance precision, generality, and computational cost, our cost model combines program analysis and machine learning: we extract program features that capture the architectural intricacies required to predict performance of GPU programs, and provide these features as input to a lightweight neural network that predicts performance. It evaluates tens of thousands of schedules per second, vs.\ seconds or minutes to compile and benchmark a single one.
Our autoscheduler then offers a spectrum of compile time -- performance tradeoffs, from a one-shot mode which relies exclusively on the cost model to schedule a program as quickly as possible, to an autotuning loop which iteratively samples and benchmarks promising candidates, fine-tuning the model as it does.

This paper contributes:
\begin{itemize}
    \item A new automatic scheduling algorithm that scales orders of magnitude better than prior work, making it possible to efficiently explore a large, rich space of GPU schedules.
    It delivers state of the art performance on a suite of real world imaging and vision pipelines, with a geomean speedup of up to \autotunespeedup\X{} over the prior state of the art GPU autoscheduler \cite{Sioutas2020},
    and competitive (\humanspeedup\X) with what the best human experts were able to achieve in an active effort to beat our automatic results.

    \item A set of schedule features that capture the architectural intricacies required to predict performance of GPU programs. 
    
    \item A GPU cost model that combines the schedule features from program analysis with machine learning. We train the model on a large set of random pipelines and a hold-one-out set of real world application programs.

\end{itemize}

\section{Why is There a Scalability Problem?} \label{sec:scalability}

\begin{figure}
    \centering
    \includegraphics[width=0.45\textwidth]{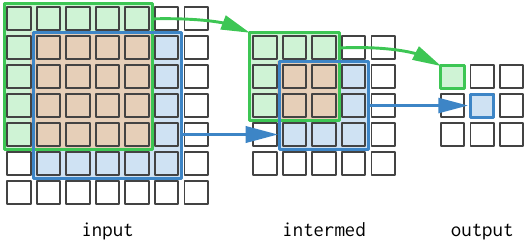}
    \caption{A small Stencil Chain with two 3\X 3 Stencils. Each point of the \code{output} accesses a 3\X 3 window of \code{intermed}, each point of which in turns accesses a 3\X 3 window of the \code{input}. Neighboring points of each stage access overlapping points of the preceding stage, which can lead to redundant recomputation or reloading of shared values, depending on the schedule. Implementing this pipeline efficiently requires balancing recompute against parallelism and memory locality when mapping to GPU resources in space and time.}
    \label{fig:stencil-chain}
\end{figure}

Consider a program consisting of a chain of stencils, where each point to be computed depends on a stencil of points from a previous stage of the program (Figure \ref{fig:stencil-chain}). This computation pattern is common in image processing algorithms, physics simulation, and deep learning architectures (e.g., a sequence of depthwise separable convolution layers). How would you implement this program for high performance on the GPU? Any implementation will need to balance the trade off between memory locality, redundant recomputation, and parallelism.

\paragraph{Separate Kernels}
The simplest approach is to schedule each stage of the chain in a separate kernel. In this case, both \code{intermed} and \code{output} will be computed in separate kernels:

\begin{lstlisting}
allocate intermed
for col, row in intermed.W, intermed.H @blocks
 compute intermed

allocate output
for col, row in output.W, output.H @blocks
 compute output
\end{lstlisting}

\noindent
Every point of \code{intermed} will be computed first, and stored in global memory, then every point of \code{output} will be computed, loading them.
As written, this implementation is inefficient. It uses parallelism at the block level but makes no use of thread parallelism.

\paragraph{Tiling the Blocks}
A more efficient implementation would \emph{tile} the blocks by splitting the outer loops into outer block loops and inner thread loops:

\begin{lstlisting}
allocate intermed
for col_o, row_o in ?, ? @blocks
 for col_i, row_i in ?, ? @threads
  compute intermed

allocate output
for col_o, row_o in ?, ? @blocks
 for col_i, row_i in ?, ? @threads
  compute output
\end{lstlisting}

\noindent
But we now have \emph{choices} for how big the tiles of each stage should be. As we consider more complex schedules, these choices will affect things like locality and redundant computation, but for now they mostly impact parallelism:
Small tiles produce many blocks, but few threads per block---plenty of parallelism across streaming multiprocessors (SMs), but potentially too little thread parallelism within each to keep it busy. Large tiles make the opposite tradeoff: less (block) parallelism across SMs, and more (threads) within each one.
16, 32, and 64 are typical choices for the innermost thread dimension, while powers of 2
are common choices for the remaining dimensions.

Importantly, by introducing block tiling, we have created $O(T^d)$ possible options to consider (where $T$ is the number of tile sizes and $d$ is the number of dimensions to tile, in this case 2). If the size of the iteration domain is small, the number of options is smaller. Larger arrays, especially if they have multiple dimensions, will have many more options. Large three- and four-dimensional tensor operations are common in image processing.
Even for this seemingly straightforward implementation, there are still many possibilities to consider.

\paragraph{Tiling the Threads}

A further optimization we can consider is tiling the thread level, as well:

\begin{lstlisting}
allocate intermed
for col_o, row_o in ?, ? @blocks
 for col_i, row_i in ?, ? @threads
  for col_ii, row_ii in ?, ? @serial, unrolled
   compute intermed

allocate output
for col_o, row_o in ?, ? @blocks
 for col_i, row_i in ?, ? @threads
  for col_ii, row_ii in ?, ? @serial, unrolled
   compute output
\end{lstlisting}

\noindent
Each thread is then responsible for computing a sub-tile's worth of points, instead of just one. This gives coarser-grained parallel tasks, and provides opportunities to exploit input reuse (fetching shared values once and reusing them in registers) and instruction-level parallelism, especially if the serial sub-tiles are \emph{unrolled}. But it also comes with tradeoffs: it reduces parallelism at the block and thread levels, and may increase the stride of memory accesses across threads, reducing bandwidth efficiency.

Typical choices for thread tiling options are small constants, and the loops are often unrolled. 
Critically, these tiling options compound with the choices at the block level, creating a space of nested decisions to consider. Attempting to exploit this \emph{nested parallelism} introduces the major challenge of enumerating schedule options for GPUs: scalability.

Even for a modest number of tiling options and a small number of dimensions, this total quickly grows into the hundreds or thousands -- and these are only options for a single stage of the program. Real applications can have 100s of stages, many with 4 or more dimensions. For example, a 3 dimension stencil chain with input size 1536x2560x8 can generate almost 3 million possible options, even for a chain of length 14. And computing each stage in its own kernel is hardly the only option we need to consider to achieve high performance.

\paragraph{Fusion Options}

The options we considered so far all exhibit poor memory locality: all intermediate values produced by \code{intermed} are computed and stored to slow global memory before any are used to compute \code{output}, at which point they have likely fallen out of cache. An alternative is to move (or fuse) the computation of \code{intermed} \emph{inside} the loop nest of \code{output}:

\begin{lstlisting}
allocate output
for col_o, row_o in ?, ? @blocks
 allocate intermed @shared_mem
 for col_i, row_i in ?, ? @threads
  for col_ii, row_ii in ?, ? @serial, unrolled
   compute intermed
 for col_i, row_i in ?, ? @threads
  for col_ii, row_ii in ?, ? @serial, unrolled
   compute output
\end{lstlisting}

\noindent
In doing so we improve memory locality, and the smaller intermediate working set can be stored in faster local memories -- shared memory if fused at the block level like here, or registers if fused one step further, all the way inside the thread level:

\begin{lstlisting}
allocate output
for col_o, row_o in ?, ? @blocks
 for col_i, row_i in ?, ? @threads
  allocate intermed @registers
  for col_ii, row_ii in ?, ? @serial, unrolled
   compute intermed
  for col_ii, row_ii in ?, ? @serial, unrolled
   compute output
\end{lstlisting}

\noindent
However, these fusion choices come at the cost of redundant recomputation of all the points in \code{intermed} where the the stencil needed by \code{output} overlaps from one tile to the next (the orange region in Figure~\ref{fig:stencil-chain}).
Fusion introduces additional choices for the level in the loop nest at which to fuse each stage, and additional tiling options within those fused blocks, further exacerbating the scalability problem.

\subsection{The Cost of Evaluating an Option}
The space of choices we have to consider to optimize a program like this for the GPU is large, rapidly reaching into the millions for real programs.
Fully compiling and benchmarking each choice is prohibitive, as it can take tens of seconds to minutes per choice.
We therefore rely on a cheaper, but still rich \emph{cost model} (\secref{sec:cost_model}) to more quickly evaluate choices.
But even this is far from free: to accurately predict the performance of complex programs on complex hardware, it applies both a wide array of static analyses to extract performance-relevant features of a given choice, and a small deep neural network to compute the nonlinear mapping from these features to ultimate performance.
Our cost model is highly optimized, but it still often takes tens of microseconds to evaluate -- orders of magnitude more than simply enumerating a choice.
And the cost function introduces an additional scalability challenge: the number of choices to evaluate grows with the number of stages in the program $n$, but the cost of evaluating the cost model also grows with $n$.
In all, this makes it infeasible to directly optimize GPU schedules using the tree search techniques introduced by prior work \cite{Adams2019}, since autoschedule times increase substantially as programs grow.

\subsection{Limitations of Graph Partitioning} \label{sec:graph-partitioning}

Tensor compilers also face scalability challenges, given the large size of neural networks. They often address this by first partitioning large programs into smaller components that can be scheduled independently. This works well for neural networks, where individual layers have high arithmetic intensity and dense connectivity, so long-range fusion is not necessary to achieve high performance.
The imaging and vision pipelines we target, however, consist primarily of numerous relatively low arithmetic intensity operations, interconnected more sparsely, and therefore require scheduling many stages together, often making complex recompute vs.\ storage tradeoffs, to optimize locality. Heuristic graph partitioning achieves scalability only by preventing a detailed schedule search from exploring such optimizations which are essential to our domain.

We studied this issue empirically by analyzing the behavior of TVM's state of the art Ansor autoscheduler~\cite{Zheng:2020:AGH} on the stencil chain pipeline discussed earlier in this section.
When scheduling a stencil chain pipeline, the graph partitioner maximally decomposes the program into $n$ independent single stencil stages, allowing no cross-layer optimization (fusion), and causing a 2.5\X{} slowdown relative to the best known schedule.

We additionally performed the same experiments while bypassing the graph partitioner. We find that the compilation time increases exponentially with the number of layers, while the performance of the discovered schedules \emph{also} decreases exponentially. This pathology can be attributed to a heuristic optimization rule that, when triggered, inlines computation of a producer into its consumers. When applied to all producers, this rule causes exponential slowdown of both compilation and execution of the compiled program. While this design is likely sensible for optimizing neural network models, we find that many domain-specific assumptions underpinning the design of such optimization rules do not transfer readily to the domain of imaging and vision pipelines.

\section{Overview of the Autoscheduler}

Our algorithm extends Adams et al.~\cite{Adams2019}'s autoscheduler. Crucially, our newly introduced hierarchical sampling, decision freezing, and memoization allow it to effectively explore a larger space of schedules, providing significant compile time speedup compared to with them disabled. Like Adams et al.~\cite{Adams2019}'s method, our autoscheduler consists of 3 major components. First, we enumerate a large space of plausible GPU schedules for a given Halide program. Second, we featurize the schedules and provide them to a learned cost model that predicts program run time. And third, we use a variant of beam search to explore the space of possible schedules. The beam search uses our hierarchical sampling strategy to make the search space exploration scalable and is guided by the cost model to search for the best performing programs. Our algorithm supports different modes of operation. It can be used to schedule a program quickly in a one-shot fashion, without any compiling or benchmarking. It can also be used for autotuning: we can generate many possible schedules, compile and benchmark them on the target GPU and use these programs to retrain the cost model, improving its ability to accurately predict program run times. This process can be repeated as desired.

\subsection{Hierarchically Sampling the Search Space} \label{sec:randomization}

\begin{figure*}
    \centering
    \includegraphics[width=0.9\textwidth]{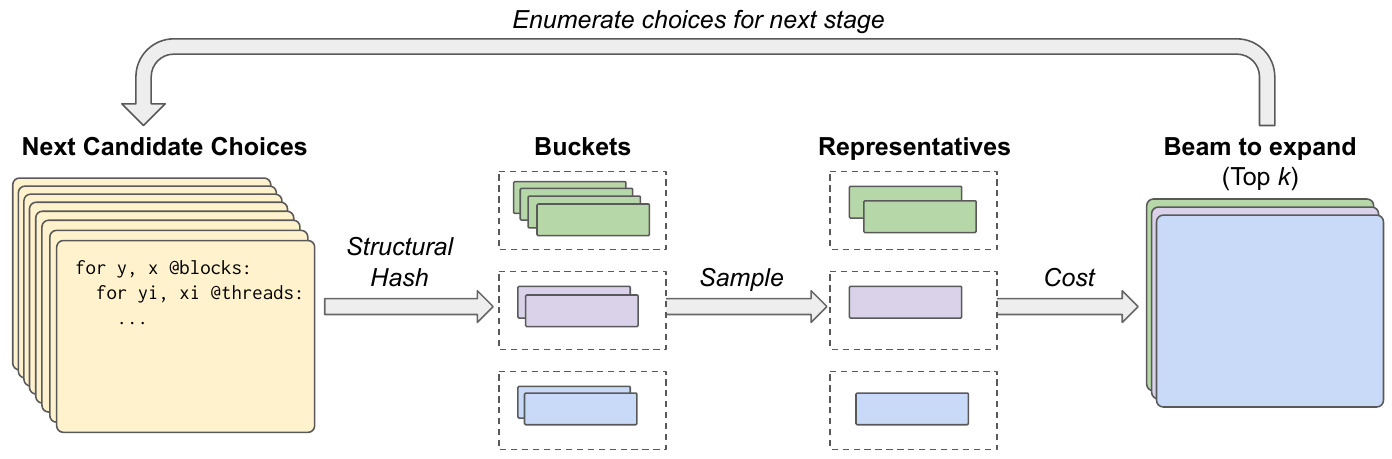}
    \caption{Our Hierarchical Sampling Strategy. A large, rich space of candidate options are enumerated. They are then grouped into buckets based on their structural similarity. We sample $log_2(B)$ representatives from each bucket. The representatives from each bucket become the final candidate states.}
    \label{fig:hierarchical-sampling}
\end{figure*}

The space of options we want to consider is too large to feasibly explore and evaluate in its entirety. How can we efficiently explore the space without featurizing and evaluating the cost model for all the possible states? Our key idea is that the search space can be partitioned into buckets based on \emph{structural similarity}, and it is sufficient to randomly select candidates from within each as \emph{representatives} to be featurized and evaluated by the cost model (Figure \ref{fig:hierarchical-sampling}). Intuitively, the representatives chosen from each group should give some sense of the expected performance of the group as a whole i.e. the expected performance of a schedule with that same structural layout. This approach stratifies the search space based on fundamental structural changes to the program's schedule. 

\begin{figure}[h]
    \centering
    \includegraphics[width=1\linewidth]{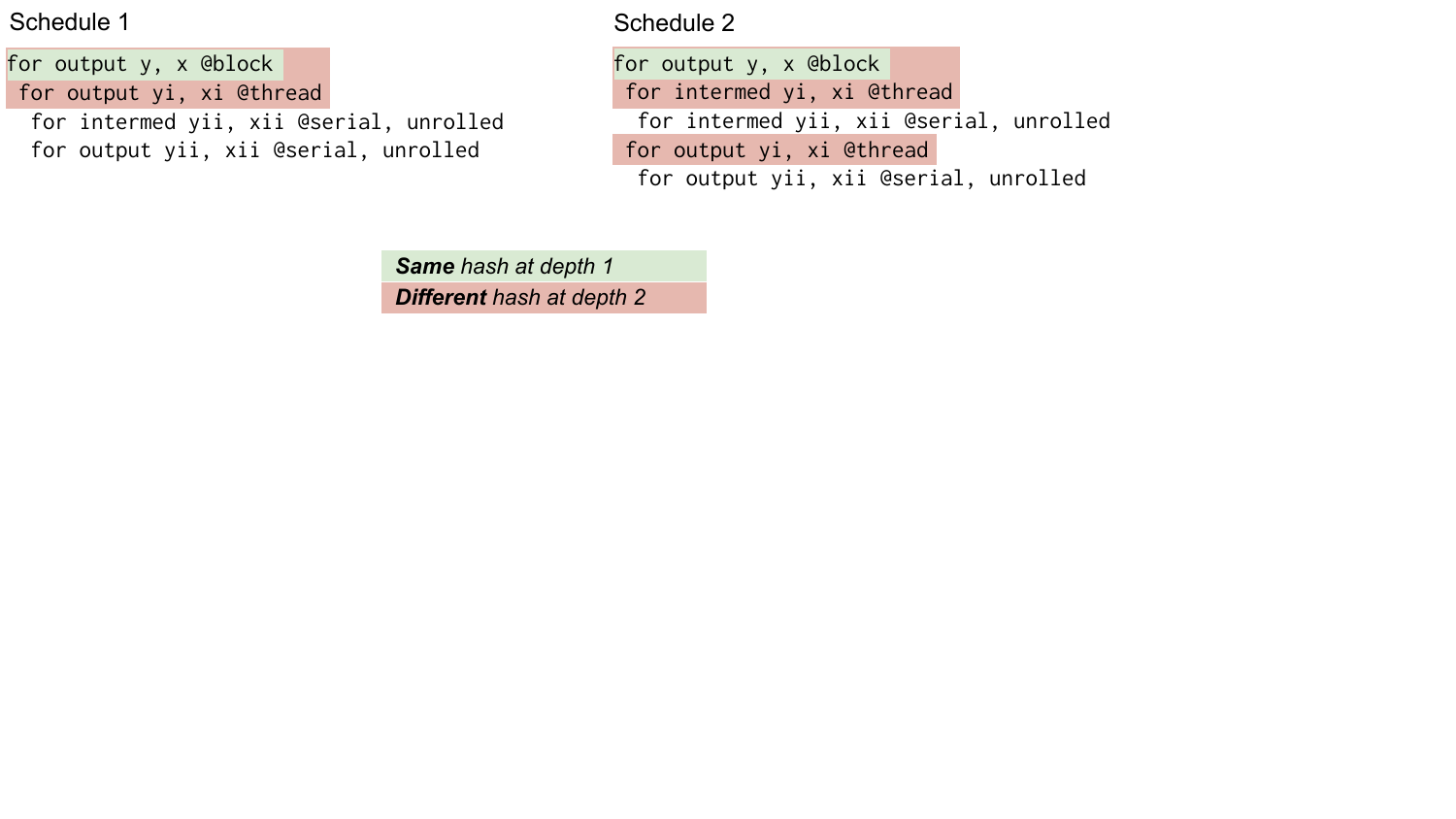}
    \caption{Structural Hashing: we hash options up to a given depth to stratify our search. Here, two different schedules have the same structure at depth 1 (which only considers block-level choices), but different structure at depth 2 (which considers both block- and thread-level choices). Equal hashes at low depth indicate at least coarse grained structural similarity. Equal hashes at high depth values indicate more fine grained structural similarity.
    }
    \label{fig:structural-hash}
\end{figure}

We compute the hash of each option up to a given depth (\figref{fig:structural-hash}), where the depth considered increases as the search process continues \cite{Adams2019} (\secref{sec:beam-search-passes}).    
For example, all options that have the same functions computed in their own kernels will have the same hash at depth 0, regardless of which other computations are fused into their kernels. But those same schedules may have different hash values at depth 1 if they have different computations at their block levels. At greater depth values, the hash function will take into account more levels of fusion: it becomes more fine grained and more buckets result. Intuitively, this allows us to control the amount of variation amongst the options in each bucket: low depth values mean few buckets where the options may only share coarse structural similarity and greater depth values mean many buckets where the options share more fine grained structural similarity. We use this to create a \emph{hierarchical} sampling process. Early on in the search, low depth values help to identify promising coarse grained structure (the compute locations for each stage of the pipeline) for candidate schedules. Then later in the search, higher depth values help to refine these coarse grained structures (the actual tile sizes to use at increasing depths).

When selecting representatives, we randomly choose only $log_2(B)$ options from each bucket (where $B$ is the number of schedules within the bucket) to be featurized and evaluated by the cost model.

\subsection{Freezing Low Cost Stages} \label{sec:freezing}

Inspired by how human experts approach scheduling by focusing their attention on the parts of the program they think will be the most costly, the second thing we do to improve scalability is to `freeze' the lowest cost stages of the program and focus our attention on the higher cost stages. During a pre-pass that only considers options that compute stages in their own kernels or inline, we enumerate options as normal, using the hierarchical sampling strategy. For the resulting schedule produced, we examine the lowest cost stages according to the cost model and `freeze' the options that were chosen for them. We then schedule the unfrozen stages without restriction. We `freeze' all but $\log_2(N)$ stages, improving the scalability for programs with many stages.

This technique serves a purpose similar to that of graph partitioning. It makes coarse grained decisions about the structure of a schedule, but it still considers all scheduling options (including fusion decisions) for the more important parts of the pipeline. In addition, instead of using heuristics, it relies on a data driven, learned approach since it is guided by the cost model.

\subsection{Memoization of Partial Schedules} \label{sec:memoization}

During the autoscheduling process, the most expensive operation performed on a candidate schedule is featurization. While these candidate schedules are unlikely to be identical, importantly, amongst them there will likely be many that exhibit some common sub-structure.

\begin{figure}[h]
    \centering
    \includegraphics[width=1\linewidth]{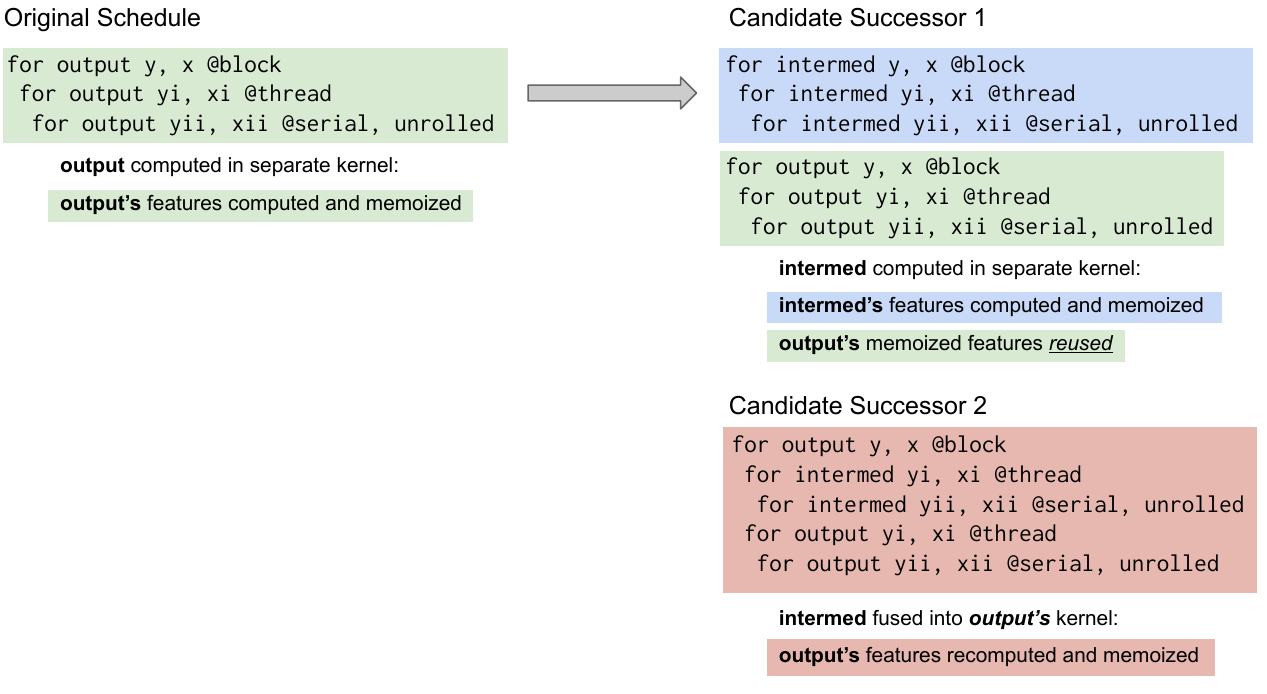}
    \caption{We memoize features on a per-kernel granularity. In the original schedule (left), \code{output} is computed in its own kernel and its features will be memoized. When generating successor schedules, \code{output}'s features can be reused if its kernel remains unchanged (Candidate Successor 1) but will be recomputed if another Func is fused into it (Candidate Successor 2).}
    \label{fig:memoization}
\end{figure}

Instead of recomputing the featurization for each sub-structure every time it occurs, we instead memoize it when it is first computed and reuse it later, amortizing its cost over all its occurrences (\figref{fig:memoization}). The features for a stage fused into its consumer's kernel are not impacted by stages fused into a different kernel so we memoize on a per-kernel granularity. Two schedules that produce the same kernel can reuse the featurization for all the stages fused into that kernel. If a new stage is fused into that particular kernel, the features for all those stages will potentially be impacted, so the featurization is at that time recomputed. As a result, schedules that include many fusion decisions will frequently need to recompute featurizations. So memoization has most impact on programs that favor schedules with separate kernels and minimal fusion.

The autoscheduler memoizes a schedule's featurization but not its cost model evaluation. Some components of the featurization change based on which other stages allocate memory outside its kernel. The set of these stages can change, so these components always need to be re-computed, even if the rest of the featurization is unchanged. And since these components will change, the featurization as a whole will also change, so a schedule's cost model evaluation also needs to be re-computed.

\section{Our Search Algorithm}

We use a variant of beam search to guide the process of enumerating options, selecting them using hierarchical sampling for evaluation by the cost model. It can be run for multiple passes, during which it uses information from previous passes to prioritize states to explore during the current pass. It maintains a priority queue of $k$ candidate states, each representing a partially scheduled loop nest. The beam search operates in 2 phases for each Func.

To help illustrate this process, we introduce a Halide pipeline based on our previous stencil chain example (simplified to use a 1D stencil across columns):

\begin{lstlisting}
Func intermed, output;
intermed(x, y) = input(x-1, y) + input(x, y) + input(x+1, y);
output(x, y) = intermed(x-1, y) + intermed(x, y) + intermed(x+1, y);
\end{lstlisting}

Halide represents this algorithm as a directed acyclic graph of \code{Func}s, where \code{output} is a consumer of producer \code{intermed}, which in turn is a consumer of \code{in}.

The search begins with a completely unscheduled pipeline and makes decisions for each \lstinline{Func} in the program in sequence, starting from the output. The schedule during this process is represented as a loop nest structure, where each level of the loop nest represents a given tiling. Every scheduling decision our algorithm makes transforms this loop nest, so we refer to this structure throughout as a natural way of describing this process. Different levels of the loop nest are labelled according to the type of parallelism they provide. The outer loops correspond to blocks. Immediately inside the outer block loops are thread loops, inside of which are serial loops.

At this point we start enumerating possible options but none of them are actually evaluated by the cost model until they are chosen by our hierarchical sampling. 

\newpage
For each \lstinline{Func} we make 2 decisions:
\begin{enumerate}
    \item Where in the currently scheduled loop nest should we compute this \lstinline{Func}?
    \item How should we tile this \lstinline{Func}?
\end{enumerate}

In the first phase of the search process, we start by making decision 1. As described in \secref{sec:scalability}, the options include computing the \lstinline{Func} in its own kernel, fusing it into one of its consumers, or inlining it, all of which introduce a tradeoff between parallelism, locality, and redundant computation.

The coarsest granularity is to compute it at the root level of the loop nest. This corresponds to launching the \lstinline{Func} in its own separate kernel. It requires no redundant recomputation but will likely exhibit poor memory locality. Its allocation will be stored in global memory, which is slow, and launching a separate kernel will incur some overhead. For \lstinline{Func output} this is the only option, since it is the output of the program.

For \lstinline{Func}s that are not outputs of the program (\lstinline{intermed}), they can also be computed at the root level but there are additional options. Each of these \lstinline{Func}s can be computed at the block level of their consumer for better memory locality, since \lstinline{output} can access a tile of \lstinline{intermed} right after they are computed.
We further place the allocation in shared memory, which is faster than global memory and L2 cache.
However, as demonstrated in Section~\ref{sec:scalability}, fusing may introduce redundant recompute.
There is also a hardware limit on the amount of shared memory available.
If a \lstinline{Func} is scheduled at this level, its loops will become thread loops.

We can also compute the \lstinline{Func} inside the thread level of its consumer. This further improves memory locality. Its allocation will be stored at the register level, which is the fastest type of memory. But this option may introduce significant redundant recompute and sacrifices parallelism since it will be computed serially by a single thread. Register memory is a very limited resource and large and/or dynamic allocations at this level may introduce costly local memory spilling.

The final option is to inline the \lstinline{Func} directly into its consumers. This option avoids storing memory altogether so exhibits the best memory locality but can easily introduce unacceptable levels of redundant recompute. If a \lstinline{Func} consists of a single stage and is called point-wise by its consumer, we always inline it. If a \lstinline{Func} is cheap to compute, it will always be considered for inlining but may be rejected later if our featurization determines that the state requires excessive recomputation.

Inline options and \lstinline{Func}s fused inside their producer's thread loop are not considered for tiling.

Next, we need to make decision 2 of tiling the functions. \lstinline{Func}s stored at the block level of their producer will be tiled immediately, since their tiling choice will have a significant impact on the featurization and cost of their producer and other siblings that are fused at the same block. For them we enumerate serial loop sizes so they become a set of thread loops outside serial loops.

\subsection{Choosing Serial Loops}
First, we enumerate inner serial loop options. These options allow loaded points to be stored in registers for faster accesses and the goal is for them to be unrolled so inputs can be reused across the unrolled loops. We do not want them to be too large because they would then reduce the amount of parallelism available at the block and thread levels and potentially increase the stride of memory accesses, which can negatively impact memory efficiency by decreasing global memory coalescing and/or increasing shared memory bank conflicts. We enumerate serial tile sizes that are small powers of 2: 1, 2, 4, and 8 in each dimension. We also consider small odd tilings (3, 5, 7) if they will enable the resulting thread loop's extent to be a multiple of the warp size (e.g. tiling an extent of 96 with a serial loop of size 3 would enable a thread loop of 32).

At this point, the compute location has been chosen and candidates scheduled at their producer's block have been tiled.

If we decide to compute the \lstinline{Func} in its own kernel, we defer its tiling to the second phase of the search to reduce the number of options to be enumerated in a single phase. At that point, it will be tiled into 3 levels. First, serial loops are chosen as above.

\subsection{Choosing Thread Loops}
After enumerating the serial loop options, we then enumerate thread loop options. Our goal in this step is to make effective use of thread parallelism while also ensuring an adequate number of blocks at the outer level. We enumerate thread loop sizes that encourage favourable warp sizes: 16, 32, 64 in the innermost loop dimension, and powers of 2 up to 16 in the other dimensions. We select as the innermost loop dimension the first dimension with extent >= 16. If there are none, we use the first dimension.

\subsection{Block Loops}
The remaining loop extents after choosing thread sizes will become the outer block loops.
A good schedule should aim to have sufficient parallelism (at least 2x the number of SMs on the GPU) to keep all the SMs busy and a balanced number of blocks that does not leave too many SMs idle.

Tiling decisions are made depending on the \lstinline{Func}'s chosen compute location. If computed at the root level, we enumerate all serial and thread loop options. If computed at the block level, we enumerate all serial loop options only: the outer loop after tiling becomes a thread loop and there is no need to tile it because it is already surrounded by a block loop. And if computed inside the thread loops, tilings are not enumerated: the resulting bounds of the Func are likely too small to make tiling worthwhile.

\subsection{Hierarchical Sampling}

At the end of each phase, once we have enumerated the search space options for a given Func, we want to featurize and evaluate them with the cost model. But as described in Section \ref{sec:scalability}, it's not feasible to  evaluate them all. Instead we apply our hierachical sampling strategy (Section \ref{sec:randomization}).

Before featurizing and evaluating the cost model, we organize all the enumerated options into buckets based on a structural hash of their loop nest. We randomly sample representatives from each bucket. These final selected options are then featurized and evaluated by the cost model and added to the beam.

\subsection{Avoiding Known Bad States} \label{sec:beam-search-passes}

The search algorithm can be run in multiple passes. During each pass, as candidate options are taken from the beam, we first compute their structural hash up to the depth dictated by the current pass. If we have previously seen that hash and the cost model informed us that it's not a promising state, we apply a cost penalty and move it back in the priority queue. Intuitively, this helps us avoid wasting time not just on the exact state under consideration, but all states that have the same hash. If we previously sampled a state as a representative during hierarchical sampling and it is evaluated poorly by the cost model, it will serve as a negative example for all the other members of its structural hash bucket and help guide us towards structural hashes that either show promise or have not yet been explored. When computing the hash, we use the pass index as the depth (\figref{fig:structural-hash}). This means that during earlier passes an equal hash value will indicate that 2 options have at least coarse grained structural similarity. During later passes, the depth increases and an equal hash value indicates more fine grained similarity. Intuitively, in the earlier passes, we explore considering only coarse structure, then in later stages start to look at more fine grained differences between options. Our results use samples with beam search (beam size 32 with 5 passes), which will consider hashes up to depth 5, as well as greedy samples (beam size 1 with 1 pass), which will only consider hashes up to depth 1, i.e. the block level.

\subsection{Pruning}

The goal when enumerating the search space is to include as many \emph{plausibly} good states as possible. We want to ensure we include serial tilings because register blocking and input reuse is an important optimization on some applications. We also want to ensure there are enough blocks to keep the SMs busy, while avoiding configurations that leave SMs idle, and enough warps to promote adequate latency hiding, while avoiding states that leave excess warp lanes idle. We prune states in the following situations:
\begin{itemize}
    \item States with excessive recompute, usually caused by inlining
    \item States that leave too many SMs idle
    \item States that exhibit poor warp lane utilization
    \item States that have serial extents that are too large to be unrolled
    \item States with allocations at the thread level that are dynamic in size or too large and likely cannot be promoted from local memory to registers
    \item States that exceed the GPU's hardware limits, including states that use too many threads or too much shared memory.
\end{itemize}

\subsection{Lowering Optimizations}

Once a pipeline is fully scheduled, it is ready to be lowered to a concrete implementation.

We apply two optimizations:
\begin{itemize}
    \item We stage producers (including ones inlined) at the thread level of their consumer. Their loaded points will be staged in an intermediate buffer, which will become register storage, allowing for faster reuse.
    \item Any serial loops that have total extent less than 16 will be unrolled.
\end{itemize}

\section{Evaluating Schedules}  \label{sec:featurization}

We design an efficient cost model for evaluating the performance of a schedule on GPU. The cost model takes a set of \emph{features} generated from a program, and feeds them into a light-weight neural network. We then train the cost model to predict the performance.

We build our cost model on top of Adams et al.'s work~\cite{Adams2019}. We compute both algorithm specific features and schedule specific features. We inherit Adams et al.'s algorithm specific features, which are histograms of various arithmetic and memory operations over the entire algorithm.

\subsection{Features}
We extend Adams et al.'s schedule specific featurization to capture important characteristics of GPU architectures. These include features for capturing how the schedule uses the different types of memory available on the GPU, how effectively it utilizes the GPU's parallelism at the block and thread levels, and the level of occupancy it achieves. In Halide, a \lstinline{Func} can have multiple update \emph{stages} that write to the same memory buffer. The features are computed for each update stage of the computation.

\paragraph{Memory access.} 

We analyze the memory access pattern by looking at the strides of the array index access with respect to loop parameters.
A suboptimal stride will typically result in poor coalescing at the global memory level or bank conflicts at the shared memory level.
We use the stride to compute the number of global memory transactions or shared memory transactions required for each stage of the pipeline.
We do this once for a representative regular warp of the stage and once again if there is a tail warp (when the loop size is not divisible by the warp size), which may exhibit different memory access behavior because its lanes might be underutilized.
These memory access counts account for amortized loads that can be reused across unrolled inner serial loops and stores that can be moved outside the block of unrolled serial loops. In addition, we have a feature for the efficiency of loads and stores at both the global and shared levels.
The efficiency is defined as the ratio of bytes used to bytes actually loaded or stored.
We also compute the memory footprint accessed at various levels of a given loop nest, including per iteration of the innermost serial loop, per thread, and per block. These footprints are delineated by memory type (global, shared, register).
In the case of global and register memory, the memory footprint gives a hint to the cost model as to expected cache behavior and register pressure.

\paragraph{Effective Parallelism}
To capture how effectively the GPU's parallelism is used, we compute its number of blocks, number of warps per block, and number of threads.
We compute each stage's warp utilization, which measures how many threads of a stage's warps are idle as a percentage of the number of threads in use across all stages computed at the block level.
A low warp utilization indicates that a stage is fused alongside another stage at the block level in a way that sacrifices thread parallelism. We also compute the number of threads that are idle as a percentage of the total number of threads made available by the hardware.

\paragraph{Occupancy}
We compute for each stage warp, block, and shared memory occupancy. We compute the ratio of maximum active warps to the maximum number of active warps hardware limit and the ratio of maximum active blocks to the maximum number of active blocks hardware limit. We also compute the ratio of shared memory to the hardware shared memory limit. We use this to compute how much block occupancy is impacted by shared memory usage. 

A complete list of our features is available in Appendix A. 

\subsection{Cost Model} \label{sec:cost_model}

Once we have the features for each stage in the computation, we feed the features for each stage into a small neural network to predict a vector of coefficients. We then use these coefficients along with our features in a cost model to predict the performance per stage and sum over all stages. We inherit our cost model design from Adams et al.'s autoscheduler~\cite{Adams2019}, but with more GPU-specific features and cost model components. 

Our network accepts as input the algorithm-specific and schedule-specific features, takes the logarithm of the schedule-specific features to compress the dynamic range, and feeds both of them into a fully-connected network to produce two embedding vectors. These two embeddings are then stacked together and passed into a fully-connected network to produce a vector of positive weights.

These features and weights are then used for computing the following costs: \emph{compute}, \emph{load}, \emph{store}, \emph{parallelism} and \emph{working set}. The compute cost accounts for the number of points computed and how effectively the warp lanes and SMs are utilized to compute them. The load and store costs differentiate between the different types of memory (global, shared, register). They take into account memory access patterns and the footprints of memory loaded and stored at various levels of the loop nests. The parallelism cost estimates the number of kernels and blocks launched. The working set cost captures register pressure and cache usage. The learned coefficients are applied as weights to each of these cost components.
Full details are available in Appendix B.

\subsection{Training Procedure}

We train our cost model on a combination of random pipelines constructed from common image processing and machine learning operation building blocks \cite{Adams2019}. Additionally, for each app in our test suite we train in a hold-one-out fashion: every \emph{other} app contributes samples to the target app's training set.

\section{Results} \label{sec:results}

\begin{figure}
\ \ \emph{\small One Shot:}\\
\includegraphics[width=0.6\linewidth]{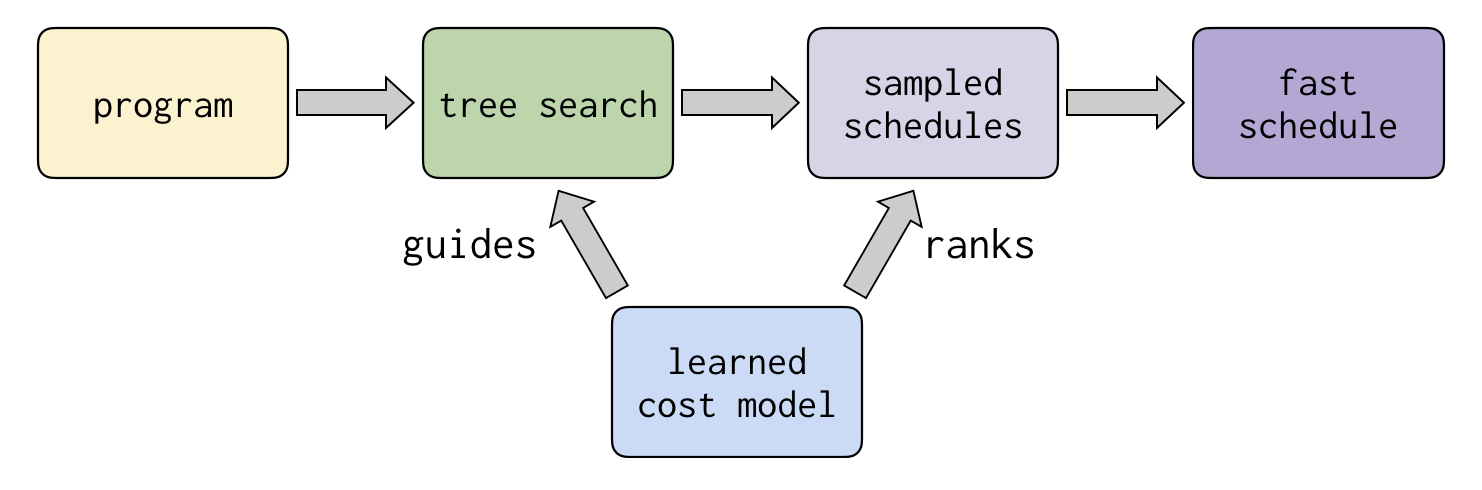}

\ \ \emph{\small Top 5:}\\
\includegraphics[width=0.8\linewidth]{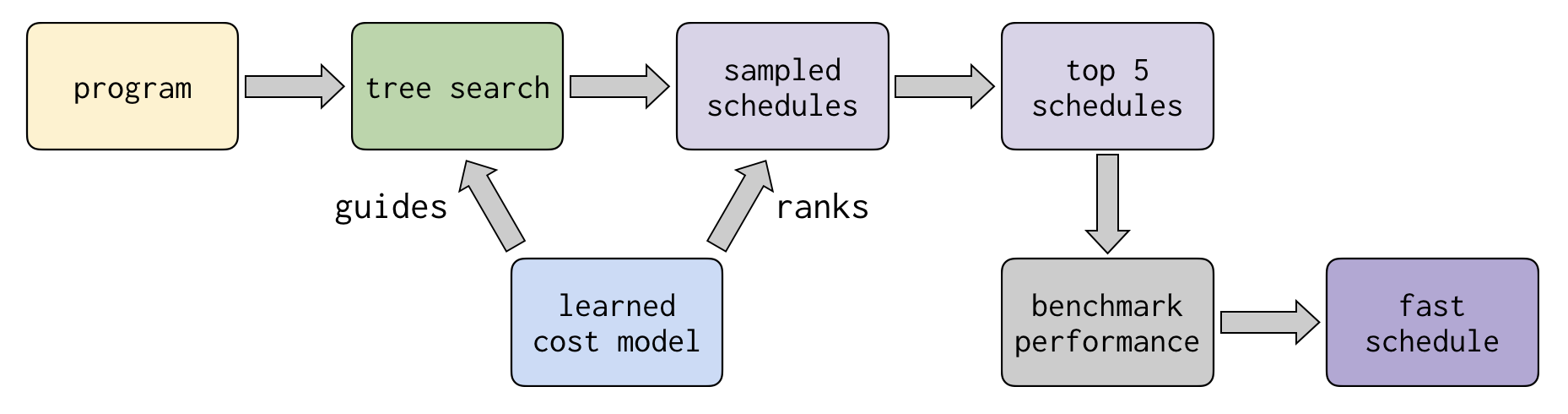}

\ \ \emph{\small Autotuned:}\\
\includegraphics[width=0.6\linewidth]{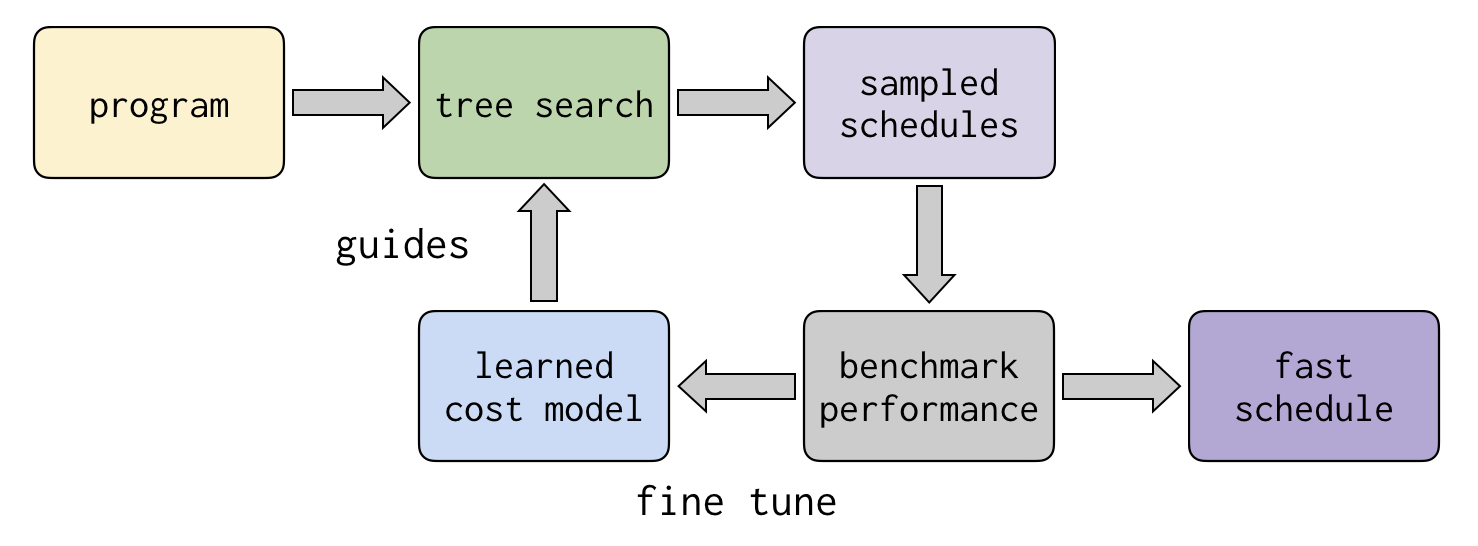}
\caption{We test our autoscheduler in three different modes, which trade increased compile time and the need to take ground-truth benchmarks for increased performance. \emph{One Shot} uses the cost model alone to rank choices. \emph{Top 5} compiles and benchmarks the top 5 choices as ranked by the cost model. \emph{Autotuned} iteratively compiles and benchmarks sampled programs, fine-tuning the cost model to the specific application as it goes.}
\label{fig:compile_modes}
\end{figure}

\begin{figure*}
    \centering
    \includegraphics[width=0.8\textwidth]{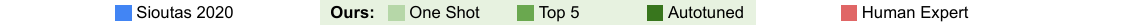}
    \includegraphics[width=1\textwidth]{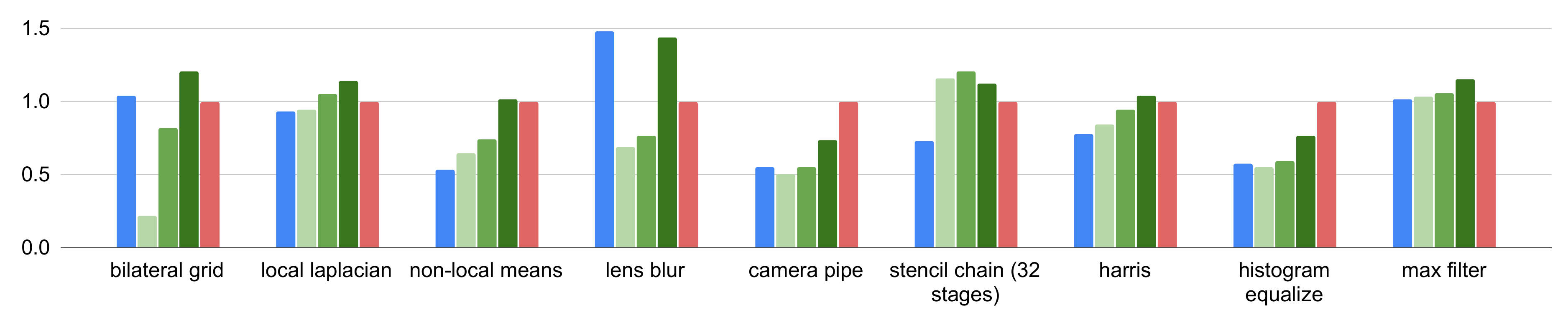}
    \includegraphics[width=1\textwidth]{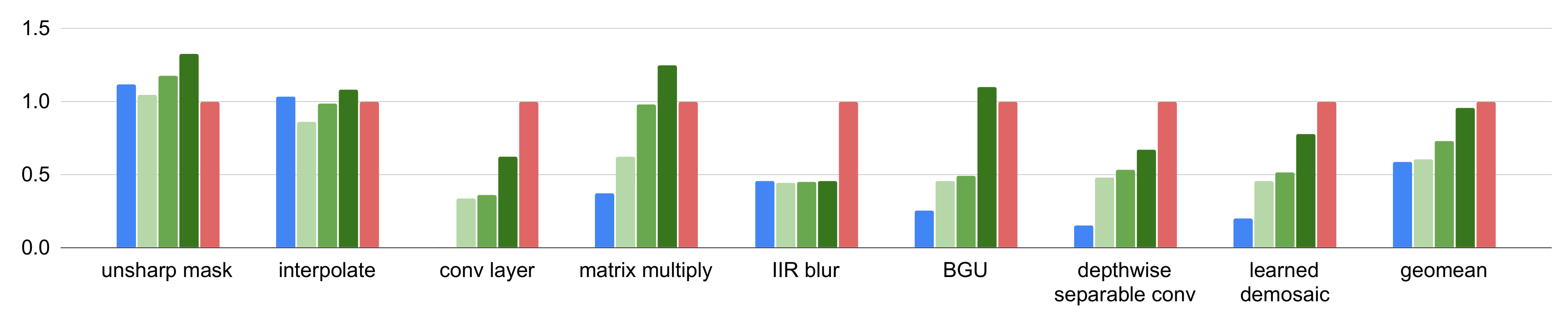}
    \caption{Throughput of our technique at three levels of compile time budget (from one shot with no benchmarking, to autotuning 1600 samples) relative to the prior state-of-the-art GPU autoscheduler~\cite{Sioutas2020} and highly tuned human expert schedules. The benchmarks are a super-set of those in prior work~\cite{Adams2019,Sioutas2020}, and span a diverse set of imaging and learning programs from a few to around one hundred stages. For the One Shot and Top 5 modes, the results presented are the median time over 100 independent trials. In all modes, our technique outperforms the prior state-of-the-art on average, and with full autotuning it matches the best human experts. Some of the human expert schedules (IIR blur, conv layer, depthwise separable conv) use scheduling options that are outside the autoscheduler's search space.
    }
    \label{fig:results-bar-chart}
\end{figure*}

We evaluate our autoscheduler on a diverse set of 17 imaging and vision programs (\figref{fig:results-bar-chart}), 
including 15 applications from the Halide repository: \emph{bilateral grid}, \emph{local laplacian}, \emph{non-local means}, \emph{lens blur}, \emph{camera pipe}, a 32-stage \emph{stencil chain}, \emph{Harris corner detection}, \emph{histogram equalize}, \emph{max filter}, \emph{unsharp mask}, \emph{interpolate}, a neural network \emph{conv layer} with ReLU activation, \emph{SGEMM} (Single float precision General Matrix Multiply), an \emph{IIR blur}, \emph{BGU} (bilateral guided upsampling). To this we added a \emph{depthwise-separable convolution}~\cite{vanhoucke2014learning}, and a learned demosaicing algorithm.

We compare our autoscheduler against the best Halide schedules experts are capable of writing by hand, using the entire scheduling language. Our experts iteratively improved these schedules during the course of this work using the best runtimes found by the autoscheduler as a target to beat (but without looking at the generated schedules), so they represent a very high bar. In many cases these human schedules substantially improve on the ones found in the Halide repository, which were used as-is by prior work. We also compare to the best existing Halide GPU autoscheduler from Sioutas et al.~\cite{Sioutas2020}. We use our technique in 3 different modes of operation, which trade off compile time and the ability to benchmark for quality of results:
\paragraph{One Shot} For each application, we generate 80 samples using a cost model trained on random pipelines and all applications beside the one being tested. We take the schedule ranked best by the cost model. No benchmarking is involved in selecting the schedule. This approach is directly comparable to Sioutas et al.~\cite{Sioutas2020}'s autoscheduler, and can run in seconds.
\paragraph{Top 5} Same as One Shot, but we consider the top 5 schedules according to the cost model, compile and benchmark them, and take the best. This is more representative of what a human might do -- construct and benchmark several promising candidates -- and it takes on the order of a minute.
\paragraph{Autotuning} We tune each application for 20 iterations, with 80 samples per iteration. All 80 samples are compiled, benchmarked, and then used to retrain the model. This mode starts from random weights, so it does not benefit from transfer learning from other applications or random pipelines, and is in fact slower than the above two modes on one application. A total of 1600 samples are generated for each application. We take the fastest schedule found during this process. This takes from 9 minutes to 71 minutes, depending on the program (Table \ref{fig:results-compile-times-sioutas}).

In all 3 modes of operation, we generate batches with 1 beam search sample (beam size = 32, 5 passes) and 79 greedy samples (beam size = 1, 1 pass).

\subsection{Post-Compile Filtering}

For the One Shot and Top 5 cases, we apply additional post-compile filtering: any samples that spill registers to local memory are removed from consideration. This performance cliff is hard to predict pre-compilation, because it depends on the vagaries of the underlying PTX compiler, which issues a warning when this happens. In one case (lens blur), all samples experienced register spilling. In this case we took the best sample that was within 50\% of the least spilling.

All our results were generated on an IBM AC922 with 2\X 20 core Power9 CPUs and 4\X NVIDIA V100 SXM2 cards with 32GB of memory. While this is an unusual processor, in no case was the time spent on the CPU a significant fraction of total runtime. All benchmarks were performed on a single V100 in isolation.

\begin{table*}
    \centering
    \caption{The number of states evaluated by the autoscheduler during a beam search, for all configurations of hierarchical sampling and freezing on and off. In parentheses is the percentage of the states evaluated relative to the configuration with both hierarchical sampling and freezing off. We highlight the lowest numbers in each row in bold. With both on, the autoscheduler evaluates on average 5.19\% of the states that would have been evaluated with both off. In the best case (bilateral grid), 0.69\% of the states are evaluated.}
    \begin{tabular}{|l||c|c|c|c|}
    \hline
        \multicolumn{5}{|c|}{Number of States Evaluated (Percentage Relative to OFF / OFF) --- Lower is Better} \\
        
\hline
\textit{Sampling} & ON & ON & OFF & OFF \\
\textit{Freezing} & ON & OFF & ON & OFF \\
\hline
\hline

        bilateral grid & \textbf{4019 (0.69\%)} & 164617 (28.44\%) & 200976 (34.72\%) & 578876 (100\%) \\
        \hline
        local laplacian & \textbf{34145 (3.18\%)} & 436254 (40.67\%) & 201263 (18.76\%) & 1072715 (100\%) \\
        \hline
        non-local means & \textbf{67750 (15.22\%)} & 182066 (40.90\%) & 443744 (99.69\%) & 445128 (100\%) \\
        \hline
        lens blur & \textbf{39268 (0.83\%)} & 2222985 (46.93\%) & 1248771 (26.36\%) & 4736690 (100\%) \\
        \hline
        camera pipe & \textbf{15345 (4.87\%)} & 159561 (50.67\%) & 139640 (44.35\%) & 314883 (100\%) \\
        \hline
        stencil chain & \textbf{22384 (8.84\%)} & 209394 (82.71\%) & 227875 (90.01\%) & 253153 (100\%) \\
        \hline
        harris & \textbf{12538 (30.56\%)} & 12340 (30.07\%) & 55267 (134.69\%) & 41034 (100\%) \\
        \hline
        hist. equalize & \textbf{10360 (13.13\%)} & 30912 (39.18\%) & 45620 (57.82\%) & 78904 (100\%) \\
        \hline
        max filter & \textbf{12160 (17.32\%)} & 36707 (52.29\%) & 58098 (82.77\%) & 70196 (100\%) \\
        \hline
        unsharp mask & \textbf{2988 (12.47\%)} & 9662 (40.34\%) & 19216 (80.22\%) & 23953 (100\%) \\
        \hline
        interpolate & \textbf{17808 (2.52\%)} & 307447 (43.56\%) & 154711 (21.92\%) & 705827 (100\%) \\
        \hline
        conv layer & \textbf{7706 (1.29\%)} & 230170 (38.60\%) & 111690 (18.73\%) & 596341 (100\%) \\
        \hline
        matrix multiply & \textbf{698 (18.38\%)} & 1853 (48.79\%) & 6783 (178.59\%) & 3798 (100\%) \\
        \hline
        IIR blur & \textbf{3150 (8.26\%)} & 13464 (35.31\%) & 14974 (39.26\%) & 38136 (100\%) \\
        \hline
        BGU & \textbf{28333 (3.21\%)} & 425325 (48.20\%) & 295303 (33.46\%) & 882435 (100\%) \\
        \hline
        dep. sep. conv & \textbf{5984 (1.78\%)} & 121769 (36.31\%) & 245637 (73.25\%) & 335338 (100\%) \\
        \hline
        learned demosaic & \textbf{50488 (4.83\%)} & 353843 (33.85\%) & 1000740 (95.73\%) & 1045372 (100\%) \\
        \hline
        \textbf{geomean} & \textbf{11740 (5.19\%)} & 95092 (42.02\%) & 121008 (53.47\%) & 226311 (100\%) \\
        \hline
    \end{tabular}
    \label{fig:results-raw-states-percentage-evaluated}
\end{table*}

\compiletimestable
\compiletimespeeduptable
\begin{figure*}
    \centering
    \includegraphics[width=1\textwidth]{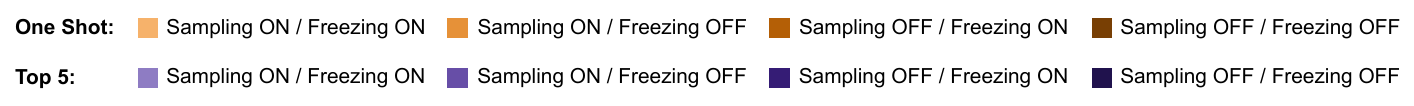}
    \includegraphics[width=1\textwidth]{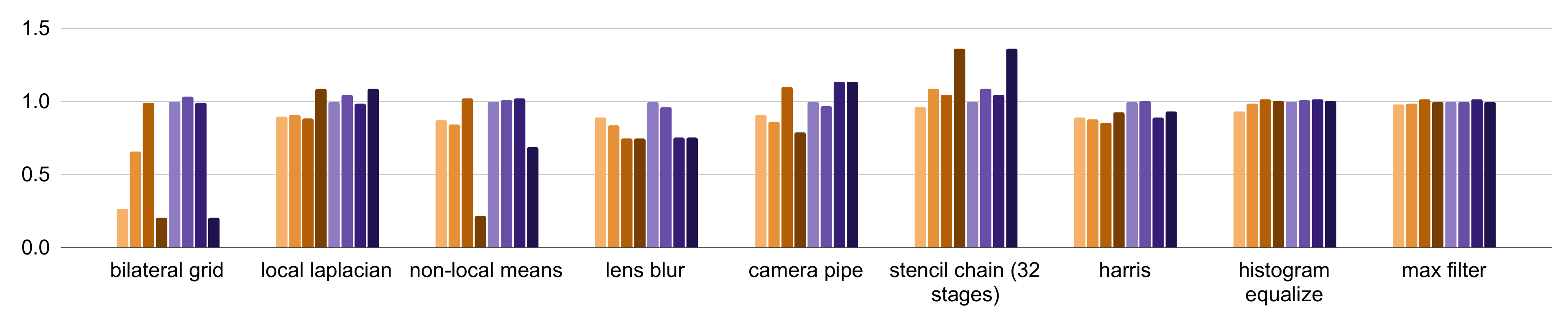}
    \includegraphics[width=1\textwidth]{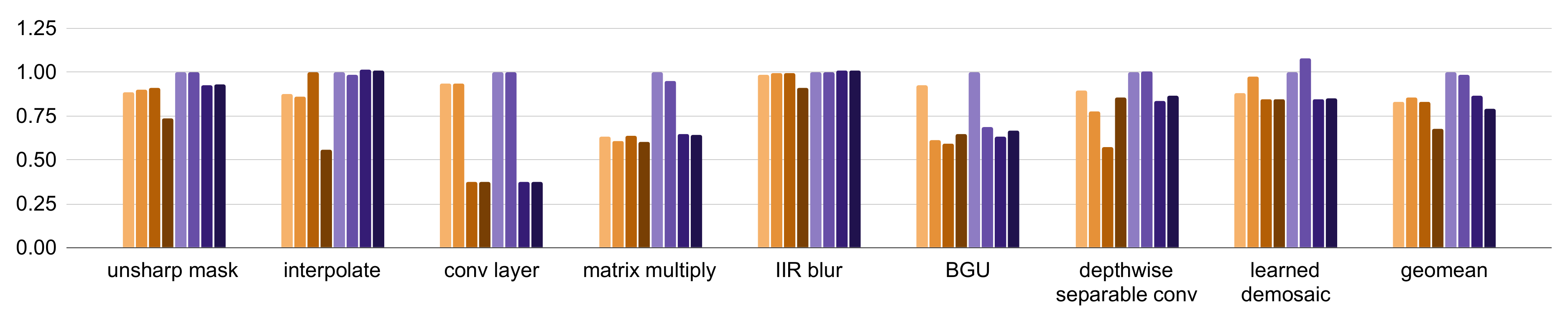}
    \caption{Our optimization strategies improve the performance of the found programs, since they stratify the search. Here we show the throughput  of our One Shot and Top 5 modes for all configurations of hierarchical sampling and freezing on and off, relative to Top 5 with sampling and freezing on. For both modes, sampling and freezing enabled on average outperforms sampling and freezing disabled. For One Shot mode, the speedup is \oneshotsamplingfreezingspeedup\Xs and for Top 5 \topfivesamplingfreezingspeedup\X. All results are the median time obtained from 100 independent trials. 
    }
    \label{fig:results-bar-chart-sampling-freezing}
\end{figure*}
\compiletimessioutas
\memorytable

\subsection{Analysis} \label{sec:analysis}

We achieve a geomean speedup over Sioutas et al.'s Halide GPU autoscheduler~\cite{Sioutas2020} of 1.07\Xs in the One Shot case, 1.30\Xs in the Top 5 case, and 1.66\Xs in the autotuning case. Our autotuned results are on par with our best known manual schedules (\humanspeedup\X).

Sioutas et al. generate code that fails to run on the \emph{conv layer}, and performs poorly on several apps, especially those outside its test set (\emph{BGU}, \emph{depthwise separable conv}, and \emph{learned demosaic}). Its single largest weakness is a search space issue with workloads that include \emph{matrix multiply} or \emph{conv layer}. A fast matrix multiplication contains an inner unrolled block over a tile of $O(MN)$ accumulators, which share the $O(M + N)$ loads required to compute each term. Sioutas et al. do not consider this form of unrolling. On \emph{BGU}, Sioutas et al. launch every step of a per-pixel 4x4 matrix solve as its own kernel. We suspect this is due to a weakness in the manually-designed cost model.

\subsection{Impact of Hierarchical Sampling, Freezing, and Memoization}

Our optimization strategies significantly reduce the compile time thanks to the reduced number of states evaluated, while making the scheduled programs faster since they strategy the search.

\subsubsection{States Evaluated}

The use of hierarchical sampling and freezing significantly reduces the number of states that are evaluated (Table \ref{fig:results-raw-states-percentage-evaluated}) during a beam search. With both sampling and freezing enabled, it evaluates on average only \meanstatesevaluated{} of the total states, and in the best case --- for bilateral grid --- only \beststatesevaluated.

\subsubsection{Compile Times}

By evaluating many fewer states, compile times are also significantly reduced (Table \ref{fig:results-compile-times}). With hierarchical sampling, freezing, and memoization enabled, compile times are reduced on average from \geomeancompiletimealloff{}s to \geomeancompiletimeallon{}s for a speedup of \geomeancompiletimespeedup\Xs (in the best case \bestcompiletimespeedup\X) (Table \ref{fig:results-compile-times-speedup}).

We also compare compile times for our method against Sioutas et al. (Table \ref{fig:results-compile-times-sioutas}).

\subsubsection{Throughput}

In Figure \ref{fig:results-bar-chart-sampling-freezing}, we compare the relative throughput in One Shot and Top 5 modes of operation for all configurations of hierarchical sampling and freezing on and off. In both modes, hierarchical sampling and freezing enabled on average outperforms hierarchical sampling and freezing disabled. For One Shot mode, the speedup is \oneshotsamplingfreezingspeedup\Xs and for Top 5 \topfivesamplingfreezingspeedup\X. This is despite the fact that with both enabled, the autoscheduler evaluates many fewer states within the search space and compiles much faster. This suggests that by stratifying the search space, it can more quickly find promising candidates. In Top 5 mode, hierarchical sampling and freezing enabled is on average the best performing configuration. In One Shot, hierarchical sampling and freezing enabled is slightly worse then sampling enabled and freezing disabled, trading a small amount of performance for faster compile times.

\subsubsection{Memory Usage}

In Table \ref{fig:results-memory-usage} we compare the memory usage of the autoscheduler with memoization both on and off. Memoization uses significant memory but reduces compile times (Table \ref{fig:results-compile-times}). We currently do not evict saved memoizations, but this functionality could be added. 

\subsection{Cost Model Evaluation}

\begin{figure*}
    \centering
    \includegraphics[width=1\textwidth]{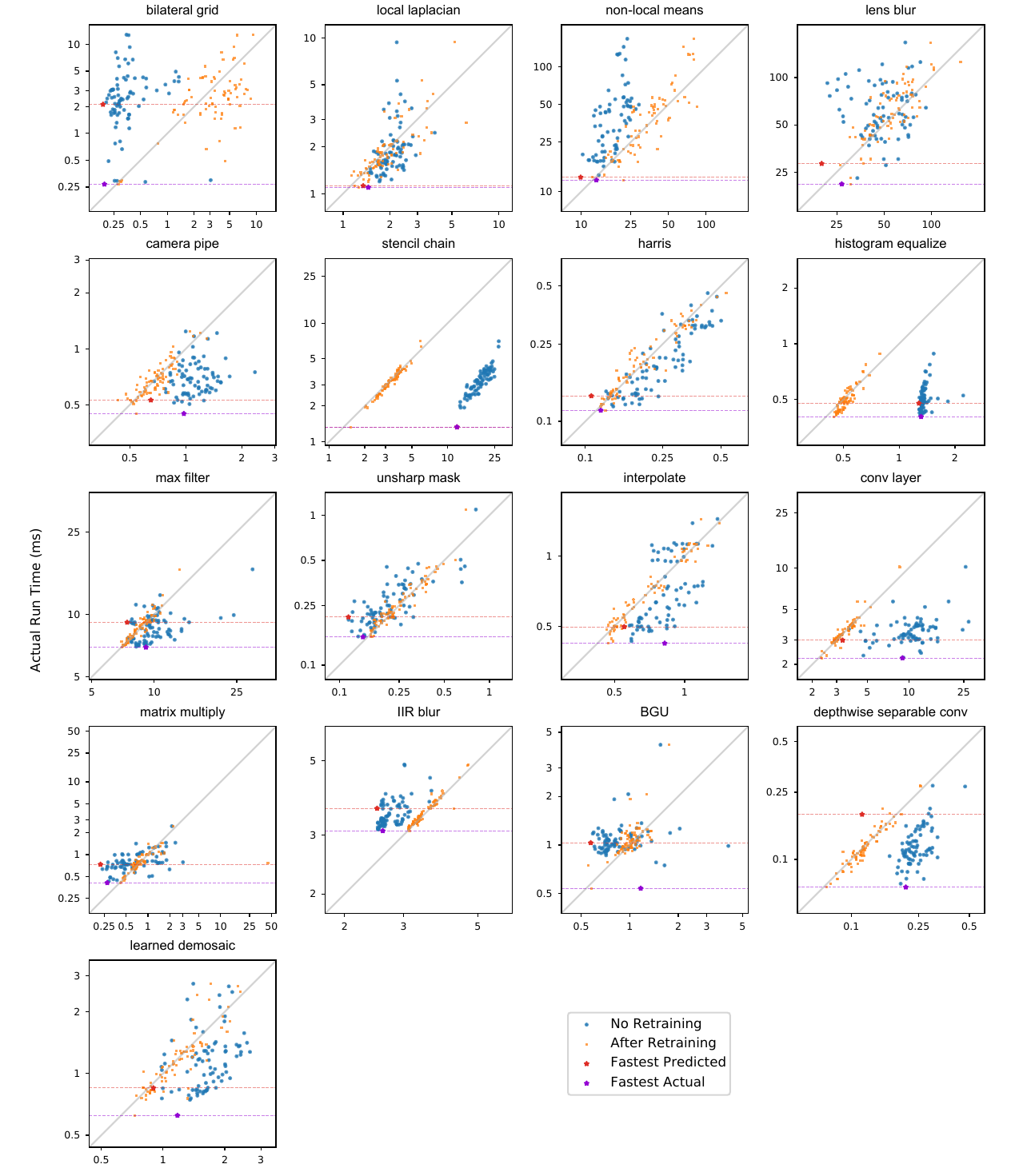}
    \includegraphics[width=1\textwidth]{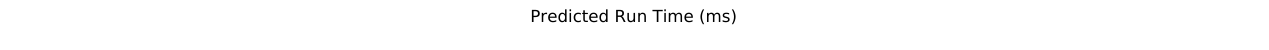}
    
    \vspace{-1em}
    
    \caption{Cost model predicted run times (x-axis) against actual run times (y-axis) for a single batch of data on each application, in two scenarios. We generate 20 batches of data for each application -- 19 are used as a training set and 1 as a test set (the points displayed in the plots). In blue, the plots show the test set predictions \textit{before} retraining. These are the predictions that would be used by the autoscheduler in One Shot mode: the cost model has never been trained on any schedules from the individual application. In orange, the plots show the test set predictions \textit{after} retraining the model on the training set, similar to how Autotuning mode works. The cost model has not been trained on the test set in either scenario. The red star in each plot is the best predicted schedule before retraining i.e. the schedule that One Shot mode will produce. The purple star is the best actual schedule. One Shot mode works best when the red star is at the bottom of the samples i.e. when it minimizes the vertical distance between it and the purple star.}
    \label{fig:results-scatter-plots}
\end{figure*}

In \figref{fig:results-scatter-plots}, we show the cost model's predictions on each application. For most of the applications, without retraining (when run in One Shot mode), the cost model is only weakly predictive of actual run times. In spite of this, the schedule selected by One Shot mode (the best schedule according to the cost model's predictions) is on average (geomean) within 48\% of the actual fastest schedule (for comparison, the average schedule in the batch is within 106\%). If we exclude bilateral grid, the selected schedule is on average within 33\% of the actual fastest schedule (the average schedule is within 84\%). This suggests that while the cost model's predictions are only weakly correlated, it makes reasonable predictions for the faster schedules in the batch. After retraining, the prediction accuracy improves significantly.
Regardless, improving the cost model's predictions before retraining is an important avenue for future work.

\subsection{Manual Schedules Outside the Search Space}

The manual schedules outperform our autoscheduler in several instances. For all of \emph{conv layer}, \emph{IIR blur}, and \emph{depthwise separable conv}, the expert schedules split and unroll the reduction loops. In \emph{BGU} and \emph{histogram equalize} the manual schedules use atomic floating point adds to memory to expose more data parallelism. In \emph{conv layer}, \emph{IIR blur}, \emph{learned demosaic}, and \emph{matrix multiply}, the manual schedule uses warp-shuffle instructions to share data between the threads in a warp without requiring a full barrier across the entire thread block. These transformations can have significant performance advantages but are not currently in the search space of our autoscheduler. Other common patterns in the manual schedules not exploited by our autoscheduler are SIMD vectorization of load instructions to reduce the total number of memory transactions, and pre-staging of stencil inputs into shared memory and/or registers to reduce the total number of loads to device memory.

Despite operating without all of these features, the autoscheduler was able to beat the manual schedules the majority of the time, and has geomean performance on par with a human expert making their best effort to beat our results. In some cases, e.g. for \emph{depthwise separable conv}, the expert spent days of focused effort re-optimizing the manual schedule after the autoscheduler's results were finalized. The large gains come in the most complex, heterogeneous applications, which are intractably difficult for humans to schedule.

\section{Related work}  \label{sec:related-work}

Earlier work on automatic array program optimization focused on affine transformations of loop nests (e.g., PLUTO~\cite{Bondhugula:2008:PAP}, Polly~\cite{Grosser:2012:PPP}, and PPCG~\cite{Verdoolaege:2013:PPC}). These works proposed that loop nests could be treated as polyhedra, and affine loop transformations could be treated as transformations on the polyhedra. This abstraction allows them to concisely express and explore many different loop optimizations, often in the form of solving an integer linear programming problem to minimize a simple cost function related to parallelism and locality. Recent work on autoscheduling \cite{tiramisu-auto} the polyhedral compiler, Tiramisu, uses an algorithm similar to ours and Adams et al. but it attempts to learn program features from loop nest representations, so its focus is different and it is CPU only.

Halide~\cite{halide2012,halide2013} takes a different approach by defining a set of domain specific rewrites to a loop nest. This allows Halide to handle a generalized form of loop fusion (called \lstinline{compute_at} and \lstinline{store_at} in Halide), which is essential, but was difficult to express in prior polyhedral optimizers. Early automatic schedulers in Halide used genetic algorithms over randomly generated schedules~\cite{halide2013, Ansel:2014:OEF}. PolyMage~\cite{Mullapudi:2015:PAO} combines ideas from the polyhedral literature and Halide, and develops an automatic scheduling technique using a heuristic cost model and a greedy stage grouping algorithm. It then compiles and benchmarks over different tile sizes. This was later extended with a richer cost model and better search algorithms~\cite{Jangda:2018:EFT}.
A parallel line of work uses a similar heuristic cost model and greedy stage grouping to handle a broader range of algorithms and schedules in Halide~\cite{Mullapudi2016, Sioutas:2019:SSH}. Sioutas et al. \cite{Sioutas2020} uses the same search algorithm but extends it to support the GPU with an expanded search space and a hand designed GPU cost model. It runs quickly without any benchmarking, but its search space is smaller compared to ours among other reasons because it only supports a single level of tiling, and as we discuss in Section \ref{sec:analysis}, this excludes a number of high performance schedules. Li et al. also developed a Halide GPU autoscheduler specialized to gradient code~\cite{Li2018}, but it only considers trivial loop fusion.

Adams et al.~\cite{Adams2019} noted that most previous approaches to automatically scheduling Halide programs focused on a restricted set of rewrites, and the heuristic cost models do not capture well the complexity of real machines. Adams et al. design a general search algorithm that can handle a broad set of scheduling options, while developing a hybrid-manual-learning-based cost model that can learn the complexity of modern hardware while remaining efficient. Unfortunately, as we discussed throughout the paper, Adams et al.'s approach does not scale well to handle the nested parallel tiling options of GPU architectures. It does not use any of our 3 scalability strategies (freezing, hierarchical sampling, memoization) and it also lacks GPU-specific features and a GPU search space and pruning techniques. Adams et al. report preliminary GPU results that are 29\%-33\% faster than a baseline \cite{Li2018} that has been superseded by Sioutas et al.'s autoscheduler~\cite{Sioutas2020}, which reports a 2\Xs improvement over it. Steiner et al.~\cite{Steiner:2021} proposed an improved cost model that includes the prediction of the unscheduled stages of the pipeline, but it is for CPU only.

Recent tensor frameworks such as TensorFlow~\cite{tensorflow2015-whitepaper} and TVM~\cite{TVM} are equipped with graph rewrite systems to optimize the composition of coarse-grained operators.
XLA~\cite{XLA:2017:XLA} applies a set of template rewrites to the computation graph using heuristic rules.
TASO~\cite{Jia:2019:TOD} takes a superoptimization approach to this problem, and generates a large collection of graph rewrites from a small set of tensor relation \emph{axioms}, and formally verifies them.
Others approach the problem using reinforcement learning agents~\cite{Paliwal:2020:RLGraph,Zhou:2020:TGO}.
These techniques focus on coarse-grained rewrites of large graphs, and not the detailed decisions of how to schedule the many dimensions \emph{within} each coarse-grained operator or operator group.

Complementary work focuses on optimizing individual tensor operators (or, equivalently, small local clusters as output by a higher-level graph rewriter).
AutoTVM~\cite{Chen:2018:Learning} automatically searches over parameters of hand-written \emph{schedule templates}, using a reinforcement learning algorithm with a statistical cost model with gradient boosted trees or tree-based recurrent neural networks. FlexTensor~\cite{Zheng:2020:FAS} and Ansor~\cite{Zheng:2020:AGH} relax the need to manually specify templates by directly enumerating them from a set of Halide-like program rewrites. Tensor Comprehensions~\cite{Vasilache:2018:TCF} employs polyhedral rewrites and autotuned tile sizes.
Because these systems focus on neural network workloads,
dominated by individual high arithmetic intensity kernels with limited opportunity for long-range fusion due to very large stencils in the channel dimensions, they are able to focus separately on small local computation graphs where scalability is less of a challenge, while relying on simpler heuristics to partition the graph~\cite{Zheng:2020:AGH}.
In contrast, we aim to schedule a broader set of programs made up of many, more diverse, and lower arithmetic intensity operations (such as the stencil chain). Fusing and jointly scheduling stages over long ranges is crucial to performance in these cases, but as described in \secref{sec:graph-partitioning}, these choices are excluded by Ansor's graph partitioner.

Related to our hierarchical sampling, Chameleon~\cite{Ahn:2020:CAC} employs k-means clustering over the tile size parameters to adaptively sample from the candidate schedules. In contrast, we perform the clustering in a hierarchical manner to adapt to the nested loop structures \secref{sec:randomization}.

Some previous work uses machine learning for compiler optimizations (e.g.,~\cite{Ashouri:2018:SCA, Haj:2020:NEV, Mendis:2019:IAP}). In contrast to most approaches in this domain, our cost model operates on a more abstract loop representation, leverages explicit program analysis and GPU architecture knowledge.

\section{Limitations \& Future Work}

Even though we consider a large space of schedules, the space of all Halide schedules is much larger still. We currently make tiling decisions on a per-\lstinline{Func} basis, but could make those decisions for all the update stages in a Halide function.
This would allow us to tile and unroll reduction variables (sequential loops that are not directly parallelizable) in update stages, which has proven an important optimization on applications like \emph{conv layer} and \emph{IIR blur}.
Making tiling decisions on a per-stage basis is likely also necessary to support parallelizing reductions (including the Halide scheduling options \lstinline{rfactor} and \lstinline{atomic}), which is an important optimization on \emph{histogram equalize}.

Register spilling often has a large performance impact on the GPU, but the spilling behavior of the downstream PTX compiler can be unpredictable. Our cost model captures factors that estimate register pressure, but can sometimes fail to predict machine-assembly-level optimizations. In our One Shot mode, we include a post-processing pass to remove programs with excessive register spilling. It would be useful to extend our cost model to predict register spilling more accurately.

\section{Conclusion}

We present a system for automatically scheduling Halide programs on GPUs that scales to a large set of scheduling options and complex pipelines. It generates code that matches experts' best effort to beat it with unconstrained manual schedules, and significantly outperforms the current state-of-the-art Halide GPU autoscheduler. We believe the key concepts of our method are likely useful for other array compilers outside of Halide.

\section*{Acknowledgments}

This work was partially funded by Toyota Research Institute, NSF awards CCF-1723445 and CCF-1846502, and DARPA agreement HR00112090017.

\appendix
\section{Featurization}\label{appendix-featurization}
We use the following GPU specific features (basic features e.g. number of productions, total points computed, etc. are reused from \cite{Adams2019}):
\begin{description}
    \item[num\_scalars] The total product of the loop extents
    \item[points\_computed\_per\_thread] The number of points of this stage computed by each thread. The product of the inner serial loops for this stage
    \item[unique\_global\_bytes\_read\_per\_realization] Number of unique bytes loaded from global memory to compute a single realization of this stage
    \item[unique\_shared\_bytes\_read\_per\_realization] Number of unique bytes loaded from shared memory to compute a single realization of this stage
    \item[unique\_register\_bytes\_read\_per\_realization] Number of unique bytes loaded from register memory to compute a single realization of this stage
    \item[unique\_global\_lines\_read\_per\_realization] Number of contiguous lines loaded from global memory to compute a single realization of this stage
    \item[unique\_shared\_lines\_read\_per\_realization] Number of contiguous lines loaded from shared memory to compute a single realization of this stage
    \item[unique\_register\_lines\_read\_per\_realization] Number of contiguous lines loaded from register memory to compute a single realization of this stage
    \item[unique\_global\_bytes\_read\_per\_thread] Number of unique bytes loaded from global memory to compute a single thread of this stage
    \item[unique\_shared\_bytes\_read\_per\_thread] Number of unique bytes loaded from shared memory to compute a single thread of this stage
    \item[unique\_register\_bytes\_read\_per\_thread] Number of unique bytes loaded from register memory to compute a single thread of this stage
    \item[unique\_global\_lines\_read\_per\_thread] Number of contiguous lines loaded from global memory to compute a single thread of this stage
    \item[unique\_shared\_lines\_read\_per\_thread] Number of contiguous lines loaded from shared memory to compute a single thread of this stage
    \item[unique\_register\_lines\_read\_per\_thread] Number of contiguous lines loaded from register memory to compute a single thread of this stage
    
    \item[global\_allocation\_bytes\_read\_per\_realization] Total sum of global memory allocation bytes accessed compute a single realization of this stage
    \item[shared\_allocation\_bytes\_read\_per\_realization] Total sum of shared memory allocation bytes accessed compute a single realization of this stage
    \item[register\_allocation\_bytes\_read\_per\_realization] Total sum of register memory allocation bytes accessed compute a single realization of this stage
    
    \item[global\_bytes\_at\_task] Number of bytes written by this stage to global memory per block
    \item[shared\_bytes\_at\_task] Number of bytes written by this stage to shared memory per block
    \item[register\_bytes\_at\_task] Number of bytes written by this stage to register memory per block
    \item[global\_innermost\_bytes\_at\_task] Number of bytes written by this stage to global memory per block, along the innermost storage dimension
    \item[shared\_innermost\_bytes\_at\_task] Number of bytes written by this stage to shared memory per block, along the innermost storage dimension
    \item[register\_innermost\_bytes\_at\_task] Number of bytes written by this stage to register memory per block, along the innermost storage dimension

    \item[num\_blocks] Number of blocks used when computing this stage
    \item[num\_warps\_per\_block] Total number of warps per block for this stage
    \item[num\_active\_warps\_per\_block] Number of warps per block for which this stage has at least 1 active thread
    \item[num\_threads\_per\_block] Number of threads per block that are used for computing this stage
    \item[expr\_branching] This stage's Strahler number: the minimum number of registers required to evaluate this stage's computation
    \item[block\_occupancy] Ratio of number of threads used to the hardware thread limit
    \item[warp\_lane\_utilization] The ratio of active threads used by this stage to the total number of active threads available (32 $\times$ number of active warps)
    \item[idle\_lane\_wastage] The ratio of idle threads in active warps to the hardware thread limit
    \item[num\_shared\_mem\_loads\_per\_block] Number of shared memory load transactions issued per block. Accounts for the number of bank conflicts of the access
    \item[num\_global\_mem\_loads\_per\_block] Number of global memory loads transactions issued per block. Accounts for the coalescing of the access
    \item[num\_shared\_mem\_stores\_per\_block] Number of shared memory stores transactions issued per block. Accounts for the bank conflicts of the access
    \item[num\_global\_mem\_stores\_per\_block] Number of global memory stores transactions issued per block. Accounts for the coaleascing of the access 

    \item[shared\_mem\_store\_efficiency] Ratio of bytes stored to shared memory to total bytes transferred by shared memory store transactions
    \item[shared\_mem\_load\_efficiency] Ratio of bytes needed by the stage from shared memory to total bytes transferred by shared memory load transactions
    
    \item[global\_mem\_store\_efficiency] Ratio of bytes stored to global memory to total bytes transferred by global memory store transactions
    \item[global\_mem\_load\_efficiency] Ratio of bytes needed by the stage from global memory to total bytes transferred by global memory load transactions
    
    \item[working\_set\_at\_thread] Sum of the allocation sizes at the thread level. Hint as to register pressure

    \item[shared\_mem\_occupancy] For compute\_ stages, ratio of total shared memory allocated at this stage's block level to shared memory hardware limit
    \item[shared\_mem\_block\_limit\_factor] Ratio of maximum active blocks allowable with the amount of shared memory allocated to the maximum active block hardware limit
    \item[max\_warp\_occupancy] Ratio of maximum active warps to maximum active warp hardware limit
    \item[max\_block\_occupancy] Ratio of maximum active blocks to maximum active block hardware limit
    
\end{description}

\section{Cost Model Components}

In the following $ci$ represents the $i$th coefficient predicted by the neural network.

\begin{lstlisting}
Let select(cond, t, f) = if cond then t else f;

compute_cost = select(inlined_calls == 0, 
                      num_scalars * c1,
                      num_scalars * c3);

num_threads = num_blocks * num_threads_per_block;
points_computed = num_threads *
                  points_computed_per_thread;

compute_cost += select(inlined_calls == 0, 
                (points_computed * c19), 
                (points_computed * c4));
                
idle_core_wastage = ceil(num_tasks / num_cores) 
                    / max(1, tasks_per_core);
                    
compute_cost *= idle_core_wastage;
compute_cost /= select(inlined_calls == 0, 
                1 - idle_lane_wastage, 1.f);

load_cost = num_realizations *
    (c5 * unique_global_lines_read_per_realization
  + c16 * unique_shared_lines_read_per_realization
  + c8 * unique_register_lines_read_per_realization
  + c6 * unique_global_bytes_read_per_realization
  + c20 * unique_shared_bytes_read_per_realization
  + c7 * unique_register_bytes_read_per_realization
  + c18 * unique_global_lines_read_per_thread
  + c17 * unique_shared_lines_read_per_thread
  + c2 * unique_register_lines_read_per_thread
  + c13 * unique_global_bytes_read_per_thread
  + c11 * unique_shared_bytes_read_per_thread
  + c0 * unique_register_bytes_read_per_thread)
  + c10 * num_scalars * unique_bytes_read_per_point
  + c12 * num_scalars * unique_lines_read_per_point 
  + c14 * num_tasks * unique_bytes_read_per_task
  + c15 * num_tasks * unique_lines_read_per_task;
  
global_mem_load_cost = num_blocks * 
    num_global_mem_loads_per_block;
global_mem_load_cost *= select(inlined_calls == 0, 
    1.f / global_mem_load_efficiency, 1);
shared_mem_load_cost = num_blocks * 
    num_shared_mem_loads_per_block;
shared_mem_load_cost *= select(inlined_calls == 0, 
    1.f / shared_mem_load_efficiency, 1);
    
load_cost += global_mem_load_cost 
          + shared_mem_load_cost;
          
shared_mem_store_cost = c29 * num_blocks * 
    num_shared_mem_stores_per_block;
global_mem_store_cost = c21 * num_blocks * 
    num_global_mem_stores_per_block;
global_mem_store_cost *= select(inlined_calls == 0,
    1.f / global_mem_store_efficiency, 1);
    
store_cost = shared_mem_store_cost 
            + global_mem_store_cost;
            
cost_of_false_sharing = select(inner_parallelism > 1,
    c22 * (num_scalars) / 
    max(1, global_innermost_bytes_at_task), 0.0f);
    
store_cost += cost_of_false_sharing;

cost_of_malloc = c24 * num_realizations;

cost_of_parallel_launches = num_productions *
    select(inner_parallelism > 1, c25, 0.0f);
cost_of_parallel_tasks = num_productions * 
    (inner_parallelism - 1) * c26;
cost_of_parallelism = cost_of_parallel_tasks 
    + cost_of_parallel_launches;
    
cost_of_working_set = working_set * c9;

cost = compute_cost + store_cost + load_cost +
       cost_of_malloc + cost_of_parallelism +
       cost_of_working_set;

\end{lstlisting}

\bibliographystyle{ACM-Reference-Format}
\bibliography{paper}


\begin{thebibliography}{30}

%%% the Web address.

\ifx \showCODEN    \undefined \def \showCODEN     #1{\unskip}     \fi
\ifx \showDOI      \undefined \def \showDOI       #1{#1}\fi
\ifx \showISBNx    \undefined \def \showISBNx     #1{\unskip}     \fi
\ifx \showISBNxiii \undefined \def \showISBNxiii  #1{\unskip}     \fi
\ifx \showISSN     \undefined \def \showISSN      #1{\unskip}     \fi
\ifx \showLCCN     \undefined \def \showLCCN      #1{\unskip}     \fi
\ifx \shownote     \undefined \def \shownote      #1{#1}          \fi
\ifx \showarticletitle \undefined \def \showarticletitle #1{#1}   \fi
\ifx \showURL      \undefined \def \showURL       {\relax}        \fi
% invisible to TeX
\providecommand\bibfield[2]{#2}
\providecommand\bibinfo[2]{#2}
\providecommand\natexlab[1]{#1}
\providecommand\showeprint[2][]{arXiv:#2}

\bibitem[\protect\citeauthoryear{Abadi, Agarwal, Barham, Brevdo, Chen, Citro,
  Corrado, Davis, Dean, Devin, Ghemawat, Goodfellow, Harp, Irving, Isard, Jia,
  Jozefowicz, Kaiser, Kudlur, Levenberg, Man\'{e}, Monga, Moore, Murray, Olah,
  Schuster, Shlens, Steiner, Sutskever, Talwar, Tucker, Vanhoucke, Vasudevan,
  Vi\'{e}gas, Vinyals, Warden, Wattenberg, Wicke, Yu, and Zheng}{Abadi
  et~al\mbox{.}}{2015}]%
        {tensorflow2015-whitepaper}
\bibfield{author}{\bibinfo{person}{Mart\'{\i}n Abadi}, \bibinfo{person}{Ashish
  Agarwal}, \bibinfo{person}{Paul Barham}, \bibinfo{person}{Eugene Brevdo},
  \bibinfo{person}{Zhifeng Chen}, \bibinfo{person}{Craig Citro},
  \bibinfo{person}{Greg~S. Corrado}, \bibinfo{person}{Andy Davis},
  \bibinfo{person}{Jeffrey Dean}, \bibinfo{person}{Matthieu Devin},
  \bibinfo{person}{Sanjay Ghemawat}, \bibinfo{person}{Ian Goodfellow},
  \bibinfo{person}{Andrew Harp}, \bibinfo{person}{Geoffrey Irving},
  \bibinfo{person}{Michael Isard}, \bibinfo{person}{Yangqing Jia},
  \bibinfo{person}{Rafal Jozefowicz}, \bibinfo{person}{Lukasz Kaiser},
  \bibinfo{person}{Manjunath Kudlur}, \bibinfo{person}{Josh Levenberg},
  \bibinfo{person}{Dandelion Man\'{e}}, \bibinfo{person}{Rajat Monga},
  \bibinfo{person}{Sherry Moore}, \bibinfo{person}{Derek Murray},
  \bibinfo{person}{Chris Olah}, \bibinfo{person}{Mike Schuster},
  \bibinfo{person}{Jonathon Shlens}, \bibinfo{person}{Benoit Steiner},
  \bibinfo{person}{Ilya Sutskever}, \bibinfo{person}{Kunal Talwar},
  \bibinfo{person}{Paul Tucker}, \bibinfo{person}{Vincent Vanhoucke},
  \bibinfo{person}{Vijay Vasudevan}, \bibinfo{person}{Fernanda Vi\'{e}gas},
  \bibinfo{person}{Oriol Vinyals}, \bibinfo{person}{Pete Warden},
  \bibinfo{person}{Martin Wattenberg}, \bibinfo{person}{Martin Wicke},
  \bibinfo{person}{Yuan Yu}, {and} \bibinfo{person}{Xiaoqiang Zheng}.}
  \bibinfo{year}{2015}\natexlab{}.
\newblock \bibinfo{title}{{TensorFlow}: Large-Scale Machine Learning on
  Heterogeneous Systems}.
\newblock
\newblock
\urldef\tempurl%
\url{https://www.tensorflow.org/}
\showURL{%
\tempurl}
\newblock
\shownote{Software available from tensorflow.org.}


\bibitem[\protect\citeauthoryear{Adams, Ma, Anderson, Baghdadi, Li, Gharbi,
  Steiner, Johnson, Fatahalian, Durand, and Ragan-Kelley}{Adams
  et~al\mbox{.}}{2019}]%
        {Adams2019}
\bibfield{author}{\bibinfo{person}{Andrew Adams}, \bibinfo{person}{Karima Ma},
  \bibinfo{person}{Luke Anderson}, \bibinfo{person}{Riyadh Baghdadi},
  \bibinfo{person}{Tzu-Mao Li}, \bibinfo{person}{Micha\"{e}l Gharbi},
  \bibinfo{person}{Benoit Steiner}, \bibinfo{person}{Steven Johnson},
  \bibinfo{person}{Kayvon Fatahalian}, \bibinfo{person}{Fr\'{e}do Durand},
  {and} \bibinfo{person}{Jonathan Ragan-Kelley}.}
  \bibinfo{year}{2019}\natexlab{}.
\newblock \showarticletitle{Learning to Optimize Halide with Tree Search and
  Random Programs}.
\newblock \bibinfo{journal}{\emph{ACM Trans. Graph. (Proc. SIGGRAPH)}}
  \bibinfo{volume}{38}, \bibinfo{number}{4}, Article \bibinfo{articleno}{121}
  (\bibinfo{date}{July} \bibinfo{year}{2019}), \bibinfo{numpages}{12}~pages.
\newblock
\showISSN{0730-0301}
\urldef\tempurl%
\url{https://doi.org/10.1145/3306346.3322967}
\showDOI{\tempurl}


\bibitem[\protect\citeauthoryear{Ahn, Pilligundla, Yazdanbakhsh, and
  Esmaeilzadeh}{Ahn et~al\mbox{.}}{2020}]%
        {Ahn:2020:CAC}
\bibfield{author}{\bibinfo{person}{Byung~Hoon Ahn}, \bibinfo{person}{Prannoy
  Pilligundla}, \bibinfo{person}{Amir Yazdanbakhsh}, {and}
  \bibinfo{person}{Hadi Esmaeilzadeh}.} \bibinfo{year}{2020}\natexlab{}.
\newblock \showarticletitle{Chameleon: Adaptive Code Optimization for Expedited
  Deep Neural Network Compilation}. In \bibinfo{booktitle}{\emph{International
  Conference on Learning Representations}}.
\newblock


\bibitem[\protect\citeauthoryear{Ansel, Kamil, Veeramachaneni, Ragan-Kelley,
  Bosboom, O'Reilly, and Amarasinghe}{Ansel et~al\mbox{.}}{2014}]%
        {Ansel:2014:OEF}
\bibfield{author}{\bibinfo{person}{Jason Ansel}, \bibinfo{person}{Shoaib
  Kamil}, \bibinfo{person}{Kalyan Veeramachaneni}, \bibinfo{person}{Jonathan
  Ragan-Kelley}, \bibinfo{person}{Jeffrey Bosboom}, \bibinfo{person}{Una-May
  O'Reilly}, {and} \bibinfo{person}{Saman Amarasinghe}.}
  \bibinfo{year}{2014}\natexlab{}.
\newblock \showarticletitle{{OpenTuner}: An extensible framework for program
  autotuning}. In \bibinfo{booktitle}{\emph{Parallel Architectures and
  Compilation}}. ACM, \bibinfo{pages}{303--316}.
\newblock
\urldef\tempurl%
\url{https://doi.org/10.1145/2628071.2628092}
\showDOI{\tempurl}


\bibitem[\protect\citeauthoryear{Ashouri, Killian, Cavazos, Palermo, and
  Silvano}{Ashouri et~al\mbox{.}}{2018}]%
        {Ashouri:2018:SCA}
\bibfield{author}{\bibinfo{person}{Amir~H Ashouri}, \bibinfo{person}{William
  Killian}, \bibinfo{person}{John Cavazos}, \bibinfo{person}{Gianluca Palermo},
  {and} \bibinfo{person}{Cristina Silvano}.} \bibinfo{year}{2018}\natexlab{}.
\newblock \showarticletitle{A survey on compiler autotuning using machine
  learning}.
\newblock \bibinfo{journal}{\emph{Computing Surveys}} \bibinfo{volume}{51},
  \bibinfo{number}{5} (\bibinfo{year}{2018}), \bibinfo{pages}{96}.
\newblock


\bibitem[\protect\citeauthoryear{Baghdadi, Merouani, Leghettas, Abdous,
  Arbaoui, Benatchba, and Amarasinghe}{Baghdadi et~al\mbox{.}}{2021}]%
        {tiramisu-auto}
\bibfield{author}{\bibinfo{person}{Riyadh Baghdadi},
  \bibinfo{person}{Massinissa Merouani}, \bibinfo{person}{Mohamed-Hicham
  Leghettas}, \bibinfo{person}{Kamel Abdous}, \bibinfo{person}{Taha Arbaoui},
  \bibinfo{person}{Karima Benatchba}, {and} \bibinfo{person}{Saman
  Amarasinghe}.} \bibinfo{year}{2021}\natexlab{}.
\newblock \showarticletitle{A Deep Learning Based Cost Model for Automatic Code
  Optimization}. In \bibinfo{booktitle}{\emph{Proceedings of the Fourth
  Conference on Machine Learning and Systems}} (San Jose, CA, USA)
  \emph{(\bibinfo{series}{MLSys 2021})}.
\newblock
\urldef\tempurl%
\url{http://groups.csail.mit.edu/commit/papers/21/tiramisu_autoscheduler.pdf}
\showURL{%
\tempurl}


\bibitem[\protect\citeauthoryear{Bondhugula, Hartono, Ramanujam, and
  Sadayappan}{Bondhugula et~al\mbox{.}}{2008}]%
        {Bondhugula:2008:PAP}
\bibfield{author}{\bibinfo{person}{Uday Bondhugula}, \bibinfo{person}{Albert
  Hartono}, \bibinfo{person}{Jagannathan Ramanujam}, {and}
  \bibinfo{person}{Ponnuswamy Sadayappan}.} \bibinfo{year}{2008}\natexlab{}.
\newblock \showarticletitle{A practical automatic polyhedral parallelizer and
  locality optimizer}.
\newblock \bibinfo{journal}{\emph{SIGPLAN Not. (Proc. PLDI)}}
  \bibinfo{volume}{43}, \bibinfo{number}{6} (\bibinfo{year}{2008}),
  \bibinfo{pages}{101--113}.
\newblock
\urldef\tempurl%
\url{https://doi.org/10.1145/1379022.1375595}
\showDOI{\tempurl}


\bibitem[\protect\citeauthoryear{Chen, Moreau, Jiang, Zheng, Yan, Cowan, Shen,
  Wang, Hu, Ceze, Guestrin, and Krishnamurthy}{Chen et~al\mbox{.}}{2018a}]%
        {TVM}
\bibfield{author}{\bibinfo{person}{Tianqi Chen}, \bibinfo{person}{Thierry
  Moreau}, \bibinfo{person}{Ziheng Jiang}, \bibinfo{person}{Lianmin Zheng},
  \bibinfo{person}{Eddie Yan}, \bibinfo{person}{Meghan Cowan},
  \bibinfo{person}{Haichen Shen}, \bibinfo{person}{Leyuan Wang},
  \bibinfo{person}{Yuwei Hu}, \bibinfo{person}{Luis Ceze},
  \bibinfo{person}{Carlos Guestrin}, {and} \bibinfo{person}{Arvind
  Krishnamurthy}.} \bibinfo{year}{2018}\natexlab{a}.
\newblock \showarticletitle{{TVM}: An Automated End-to-End Optimizing Compiler
  for Deep Learning}. In \bibinfo{booktitle}{\emph{Proceedings of the 13th
  {USENIX} Conference on Operating Systems Design and Implementation}}
  (Carlsbad, CA, USA) \emph{(\bibinfo{series}{OSDI'18})}.
  \bibinfo{publisher}{{USENIX} Association}, \bibinfo{address}{USA},
  \bibinfo{pages}{579–594}.
\newblock
\showISBNx{9781931971478}


\bibitem[\protect\citeauthoryear{Chen, Zheng, Yan, Jiang, Moreau, Ceze,
  Guestrin, and Krishnamurthy}{Chen et~al\mbox{.}}{2018b}]%
        {Chen:2018:Learning}
\bibfield{author}{\bibinfo{person}{Tianqi Chen}, \bibinfo{person}{Lianmin
  Zheng}, \bibinfo{person}{Eddie Yan}, \bibinfo{person}{Ziheng Jiang},
  \bibinfo{person}{Thierry Moreau}, \bibinfo{person}{Luis Ceze},
  \bibinfo{person}{Carlos Guestrin}, {and} \bibinfo{person}{Arvind
  Krishnamurthy}.} \bibinfo{year}{2018}\natexlab{b}.
\newblock \showarticletitle{Learning to optimize tensor programs}. In
  \bibinfo{booktitle}{\emph{Advances in Neural Information Processing
  Systems}}. \bibinfo{pages}{3389--3400}.
\newblock


\bibitem[\protect\citeauthoryear{Grosser, Groesslinger, and Lengauer}{Grosser
  et~al\mbox{.}}{2012}]%
        {Grosser:2012:PPP}
\bibfield{author}{\bibinfo{person}{Tobias Grosser}, \bibinfo{person}{Armin
  Groesslinger}, {and} \bibinfo{person}{Christian Lengauer}.}
  \bibinfo{year}{2012}\natexlab{}.
\newblock \showarticletitle{Polly — performing polyhedral optimizations on a
  low-level intermediate representation}.
\newblock \bibinfo{journal}{\emph{Parallel Processing Letters}}
  \bibinfo{volume}{22}, \bibinfo{number}{04} (\bibinfo{year}{2012}),
  \bibinfo{pages}{1250010}.
\newblock
\urldef\tempurl%
\url{https://doi.org/10.1142/S0129626412500107}
\showDOI{\tempurl}


\bibitem[\protect\citeauthoryear{Haj-Ali, Ahmed, Willke, Shao, Asanovic, and
  Stoica}{Haj-Ali et~al\mbox{.}}{2020}]%
        {Haj:2020:NEV}
\bibfield{author}{\bibinfo{person}{Ameer Haj-Ali}, \bibinfo{person}{Nesreen~K
  Ahmed}, \bibinfo{person}{Ted Willke}, \bibinfo{person}{Yakun~Sophia Shao},
  \bibinfo{person}{Krste Asanovic}, {and} \bibinfo{person}{Ion Stoica}.}
  \bibinfo{year}{2020}\natexlab{}.
\newblock \showarticletitle{{NeuroVectorizer}: end-to-end vectorization with
  deep reinforcement learning}. \bibinfo{pages}{242--255}.
\newblock
\urldef\tempurl%
\url{https://doi.org/10.1145/3368826.3377928}
\showDOI{\tempurl}


\bibitem[\protect\citeauthoryear{Jangda and Bondhugula}{Jangda and
  Bondhugula}{2018}]%
        {Jangda:2018:EFT}
\bibfield{author}{\bibinfo{person}{Abhinav Jangda} {and} \bibinfo{person}{Uday
  Bondhugula}.} \bibinfo{year}{2018}\natexlab{}.
\newblock \showarticletitle{An Effective Fusion and Tile Size Model for
  Optimizing Image Processing Pipelines}.
\newblock \bibinfo{journal}{\emph{SIGPLAN Not. (Proc. PPoPP)}}
  \bibinfo{volume}{53}, \bibinfo{number}{1} (\bibinfo{date}{Feb.}
  \bibinfo{year}{2018}), \bibinfo{pages}{261–275}.
\newblock
\urldef\tempurl%
\url{https://doi.org/10.1145/3178487.3178507}
\showDOI{\tempurl}


\bibitem[\protect\citeauthoryear{Jia, Padon, Thomas, Warszawski, Zaharia, and
  Aiken}{Jia et~al\mbox{.}}{2019}]%
        {Jia:2019:TOD}
\bibfield{author}{\bibinfo{person}{Zhihao Jia}, \bibinfo{person}{Oded Padon},
  \bibinfo{person}{James Thomas}, \bibinfo{person}{Todd Warszawski},
  \bibinfo{person}{Matei Zaharia}, {and} \bibinfo{person}{Alex Aiken}.}
  \bibinfo{year}{2019}\natexlab{}.
\newblock \showarticletitle{{TASO}: Optimizing Deep Learning Computation with
  Automatic Generation of Graph Substitutions}. In
  \bibinfo{booktitle}{\emph{Proceedings of the ACM Symposium on Operating
  Systems Principles (SOSP)}}. \bibinfo{publisher}{ACM},
  \bibinfo{pages}{47–62}.
\newblock
\urldef\tempurl%
\url{https://doi.org/10.1145/3341301.3359630}
\showDOI{\tempurl}


\bibitem[\protect\citeauthoryear{Li, Gharbi, Adams, Durand, and
  Ragan-Kelley}{Li et~al\mbox{.}}{2018}]%
        {Li2018}
\bibfield{author}{\bibinfo{person}{Tzu-Mao Li}, \bibinfo{person}{Micha{\"e}l
  Gharbi}, \bibinfo{person}{Andrew Adams}, \bibinfo{person}{Fr{\'e}do Durand},
  {and} \bibinfo{person}{Jonathan Ragan-Kelley}.}
  \bibinfo{year}{2018}\natexlab{}.
\newblock \showarticletitle{Differentiable programming for image processing and
  deep learning in {Halide}}.
\newblock \bibinfo{journal}{\emph{ACM Trans. Graph. (Proc. SIGGRAPH)}}
  \bibinfo{volume}{37}, \bibinfo{number}{4} (\bibinfo{year}{2018}),
  \bibinfo{pages}{139:1--139:13}.
\newblock
\urldef\tempurl%
\url{https://doi.org/10.1145/3197517.3201383}
\showDOI{\tempurl}


\bibitem[\protect\citeauthoryear{Mendis, Renda, Amarasinghe, and Carbin}{Mendis
  et~al\mbox{.}}{2019}]%
        {Mendis:2019:IAP}
\bibfield{author}{\bibinfo{person}{Charith Mendis}, \bibinfo{person}{Alex
  Renda}, \bibinfo{person}{Saman Amarasinghe}, {and} \bibinfo{person}{Michael
  Carbin}.} \bibinfo{year}{2019}\natexlab{}.
\newblock \showarticletitle{Ithemal: Accurate, portable and fast basic block
  throughput estimation using deep neural networks}. In
  \bibinfo{booktitle}{\emph{International Conference on Machine Learning}}.
  PMLR, \bibinfo{pages}{4505--4515}.
\newblock


\bibitem[\protect\citeauthoryear{Mullapudi, Adams, Sharlet, Ragan-Kelley, and
  Fatahalian}{Mullapudi et~al\mbox{.}}{2016}]%
        {Mullapudi2016}
\bibfield{author}{\bibinfo{person}{Ravi~Teja Mullapudi},
  \bibinfo{person}{Andrew Adams}, \bibinfo{person}{Dillon Sharlet},
  \bibinfo{person}{Jonathan Ragan-Kelley}, {and} \bibinfo{person}{Kayvon
  Fatahalian}.} \bibinfo{year}{2016}\natexlab{}.
\newblock \showarticletitle{Automatically Scheduling Halide Image Processing
  Pipelines}.
\newblock \bibinfo{journal}{\emph{ACM Trans. Graph. (Proc. SIGGRAPH)}}
  \bibinfo{volume}{35}, \bibinfo{number}{4}, Article \bibinfo{articleno}{83}
  (\bibinfo{year}{2016}), \bibinfo{numpages}{11}~pages.
\newblock
\showISSN{0730-0301}
\urldef\tempurl%
\url{https://doi.org/10.1145/2897824.2925952}
\showDOI{\tempurl}


\bibitem[\protect\citeauthoryear{Mullapudi, Vasista, and Bondhugula}{Mullapudi
  et~al\mbox{.}}{2015}]%
        {Mullapudi:2015:PAO}
\bibfield{author}{\bibinfo{person}{Ravi~Teja Mullapudi}, \bibinfo{person}{Vinay
  Vasista}, {and} \bibinfo{person}{Uday Bondhugula}.}
  \bibinfo{year}{2015}\natexlab{}.
\newblock \showarticletitle{{PolyMage}: Automatic Optimization for Image
  Processing Pipelines}.
\newblock \bibinfo{journal}{\emph{SIGPLAN Not. (Proc. ASPLOS)}}
  \bibinfo{volume}{43}, \bibinfo{number}{1} (\bibinfo{year}{2015}),
  \bibinfo{pages}{429--443}.
\newblock
\urldef\tempurl%
\url{https://doi.org/10.1145/2775054.2694364}
\showDOI{\tempurl}


\bibitem[\protect\citeauthoryear{Paliwal, Gimeno, Nair, Li, Lubin, Kohli, and
  Vinyals}{Paliwal et~al\mbox{.}}{2020}]%
        {Paliwal:2020:RLGraph}
\bibfield{author}{\bibinfo{person}{Aditya Paliwal}, \bibinfo{person}{Felix
  Gimeno}, \bibinfo{person}{Vinod Nair}, \bibinfo{person}{Yujia Li},
  \bibinfo{person}{Miles Lubin}, \bibinfo{person}{Pushmeet Kohli}, {and}
  \bibinfo{person}{Oriol Vinyals}.} \bibinfo{year}{2020}\natexlab{}.
\newblock \showarticletitle{Reinforced Genetic Algorithm Learning for
  Optimizing Computation Graphs}. In \bibinfo{booktitle}{\emph{Proceedings of
  the International Conference on Learning Representations (ICLR)}}.
\newblock


\bibitem[\protect\citeauthoryear{Ragan-Kelley, Adams, Paris, Levoy,
  Amarasinghe, and Durand}{Ragan-Kelley et~al\mbox{.}}{2012}]%
        {halide2012}
\bibfield{author}{\bibinfo{person}{Jonathan Ragan-Kelley},
  \bibinfo{person}{Andrew Adams}, \bibinfo{person}{Sylvain Paris},
  \bibinfo{person}{Marc Levoy}, \bibinfo{person}{Saman Amarasinghe}, {and}
  \bibinfo{person}{Fr\'{e}do Durand}.} \bibinfo{year}{2012}\natexlab{}.
\newblock \showarticletitle{Decoupling Algorithms from Schedules for Easy
  Optimization of Image Processing Pipelines}.
\newblock \bibinfo{journal}{\emph{ACM Trans. Graph. (Proc. SIGGRAPH)}}
  \bibinfo{volume}{31}, \bibinfo{number}{4}, Article \bibinfo{articleno}{32}
  (\bibinfo{year}{2012}), \bibinfo{numpages}{12}~pages.
\newblock
\showISSN{0730-0301}
\urldef\tempurl%
\url{https://doi.org/10.1145/2185520.2185528}
\showDOI{\tempurl}


\bibitem[\protect\citeauthoryear{Ragan-Kelley, Barnes, Adams, Paris, Durand,
  and Amarasinghe}{Ragan-Kelley et~al\mbox{.}}{2013}]%
        {halide2013}
\bibfield{author}{\bibinfo{person}{Jonathan Ragan-Kelley},
  \bibinfo{person}{Connelly Barnes}, \bibinfo{person}{Andrew Adams},
  \bibinfo{person}{Sylvain Paris}, \bibinfo{person}{Fr\'{e}do Durand}, {and}
  \bibinfo{person}{Saman Amarasinghe}.} \bibinfo{year}{2013}\natexlab{}.
\newblock \showarticletitle{Halide: A Language and Compiler for Optimizing
  Parallelism, Locality, and Recomputation in Image Processing Pipelines}.
\newblock \bibinfo{journal}{\emph{SIGPLAN Not. (Proc. PLDI)}}
  \bibinfo{volume}{48}, \bibinfo{number}{6} (\bibinfo{year}{2013}),
  \bibinfo{pages}{519–530}.
\newblock
\urldef\tempurl%
\url{https://doi.org/10.1145/2491956.2462176}
\showDOI{\tempurl}


\bibitem[\protect\citeauthoryear{Sioutas, Stuijk, Basten, Corporaal, and
  Somers}{Sioutas et~al\mbox{.}}{2020}]%
        {Sioutas2020}
\bibfield{author}{\bibinfo{person}{Savvas Sioutas}, \bibinfo{person}{Sander
  Stuijk}, \bibinfo{person}{Twan Basten}, \bibinfo{person}{Henk Corporaal},
  {and} \bibinfo{person}{Lou Somers}.} \bibinfo{year}{2020}\natexlab{}.
\newblock \showarticletitle{Schedule Synthesis for Halide Pipelines on GPUs}.
\newblock \bibinfo{journal}{\emph{ACM Trans. Archit. Code Optim.}}
  \bibinfo{volume}{17}, \bibinfo{number}{3}, Article \bibinfo{articleno}{23}
  (\bibinfo{year}{2020}), \bibinfo{numpages}{25}~pages.
\newblock
\showISSN{1544-3566}
\urldef\tempurl%
\url{https://doi.org/10.1145/3406117}
\showDOI{\tempurl}


\bibitem[\protect\citeauthoryear{Sioutas, Stuijk, Waeijen, Basten, Corporaal,
  and Somers}{Sioutas et~al\mbox{.}}{2019}]%
        {Sioutas:2019:SSH}
\bibfield{author}{\bibinfo{person}{Savvas Sioutas}, \bibinfo{person}{Sander
  Stuijk}, \bibinfo{person}{Luc Waeijen}, \bibinfo{person}{Twan Basten},
  \bibinfo{person}{Henk Corporaal}, {and} \bibinfo{person}{Lou Somers}.}
  \bibinfo{year}{2019}\natexlab{}.
\newblock \showarticletitle{Schedule Synthesis for {Halide} Pipelines through
  Reuse Analysis}.
\newblock \bibinfo{journal}{\emph{Trans. Archit. Code Optim.}}
  \bibinfo{volume}{16}, \bibinfo{number}{2} (\bibinfo{year}{2019}),
  \bibinfo{pages}{10:1--10:22}.
\newblock
\urldef\tempurl%
\url{https://doi.org/10.1145/3310248}
\showDOI{\tempurl}


\bibitem[\protect\citeauthoryear{Steiner, Cummins, He, and Leather}{Steiner
  et~al\mbox{.}}{2021}]%
        {Steiner:2021}
\bibfield{author}{\bibinfo{person}{Benoit Steiner}, \bibinfo{person}{Chris
  Cummins}, \bibinfo{person}{Horace He}, {and} \bibinfo{person}{Hugh Leather}.}
  \bibinfo{year}{2021}\natexlab{}.
\newblock \showarticletitle{Value Learning for Throughput Optimization of Deep
  Learning Workloads}. In \bibinfo{booktitle}{\emph{Proceedings of Machine
  Learning and Systems}}.
\newblock


\bibitem[\protect\citeauthoryear{Team}{Team}{2017}]%
        {XLA:2017:XLA}
\bibfield{author}{\bibinfo{person}{The~{XLA} Team}.}
  \bibinfo{year}{2017}\natexlab{}.
\newblock \bibinfo{title}{{XLA} -- {TensorFlow} compiled}.
\newblock
  \bibinfo{howpublished}{\url{https://developers.googleblog.com/2017/03/xla-tensorflow-compiled.html}}.
\newblock
\newblock
\shownote{Accessed: 2020-08-19.}


\bibitem[\protect\citeauthoryear{Vanhoucke}{Vanhoucke}{2014}]%
        {vanhoucke2014learning}
\bibfield{author}{\bibinfo{person}{Vincent Vanhoucke}.}
  \bibinfo{year}{2014}\natexlab{}.
\newblock \showarticletitle{Learning visual representations at scale}.
\newblock \bibinfo{journal}{\emph{ICLR invited talk}}  \bibinfo{volume}{1}
  (\bibinfo{year}{2014}), \bibinfo{pages}{2}.
\newblock


\bibitem[\protect\citeauthoryear{Vasilache, Zinenko, Theodoridis, Goyal,
  DeVito, Moses, Verdoolaege, Adams, and Cohen}{Vasilache
  et~al\mbox{.}}{2018}]%
        {Vasilache:2018:TCF}
\bibfield{author}{\bibinfo{person}{Nicolas Vasilache},
  \bibinfo{person}{Oleksandr Zinenko}, \bibinfo{person}{Theodoros Theodoridis},
  \bibinfo{person}{Priya Goyal}, \bibinfo{person}{Zachary DeVito},
  \bibinfo{person}{William~S Moses}, \bibinfo{person}{Sven Verdoolaege},
  \bibinfo{person}{Andrew Adams}, {and} \bibinfo{person}{Albert Cohen}.}
  \bibinfo{year}{2018}\natexlab{}.
\newblock \showarticletitle{Tensor comprehensions: Framework-agnostic
  high-performance machine learning abstractions}.
\newblock \bibinfo{journal}{\emph{arXiv:1802.04730}} (\bibinfo{year}{2018}).
\newblock


\bibitem[\protect\citeauthoryear{Verdoolaege, Carlos~Juega, Cohen,
  Ignacio~G\'{o}mez, Tenllado, and Catthoor}{Verdoolaege et~al\mbox{.}}{2013}]%
        {Verdoolaege:2013:PPC}
\bibfield{author}{\bibinfo{person}{Sven Verdoolaege}, \bibinfo{person}{Juan
  Carlos~Juega}, \bibinfo{person}{Albert Cohen}, \bibinfo{person}{Jos\'{e}
  Ignacio~G\'{o}mez}, \bibinfo{person}{Christian Tenllado}, {and}
  \bibinfo{person}{Francky Catthoor}.} \bibinfo{year}{2013}\natexlab{}.
\newblock \showarticletitle{Polyhedral Parallel Code Generation for CUDA}.
\newblock \bibinfo{journal}{\emph{Trans. Archit. Code Optim.}}, Article
  \bibinfo{articleno}{54} (\bibinfo{year}{2013}), \bibinfo{numpages}{23}~pages.
\newblock
\urldef\tempurl%
\url{https://doi.org/10.1145/2400682.2400713}
\showDOI{\tempurl}


\bibitem[\protect\citeauthoryear{Zheng, Jia, Sun, Wu, Yu, Haj-Ali, Wang, Yang,
  Zhuo, Sen, et~al\mbox{.}}{Zheng et~al\mbox{.}}{2020a}]%
        {Zheng:2020:AGH}
\bibfield{author}{\bibinfo{person}{Lianmin Zheng}, \bibinfo{person}{Chengfan
  Jia}, \bibinfo{person}{Minmin Sun}, \bibinfo{person}{Zhao Wu},
  \bibinfo{person}{Cody~Hao Yu}, \bibinfo{person}{Ameer Haj-Ali},
  \bibinfo{person}{Yida Wang}, \bibinfo{person}{Jun Yang},
  \bibinfo{person}{Danyang Zhuo}, \bibinfo{person}{Koushik Sen},
  {et~al\mbox{.}}} \bibinfo{year}{2020}\natexlab{a}.
\newblock \showarticletitle{Ansor: Generating High-Performance Tensor Programs
  for Deep Learning}.
\newblock \bibinfo{journal}{\emph{arXiv preprint arXiv:2006.06762}}
  (\bibinfo{year}{2020}).
\newblock


\bibitem[\protect\citeauthoryear{Zheng, Liang, Wang, Chen, and Sheng}{Zheng
  et~al\mbox{.}}{2020b}]%
        {Zheng:2020:FAS}
\bibfield{author}{\bibinfo{person}{Size Zheng}, \bibinfo{person}{Yun Liang},
  \bibinfo{person}{Shuo Wang}, \bibinfo{person}{Renze Chen}, {and}
  \bibinfo{person}{Kaiwen Sheng}.} \bibinfo{year}{2020}\natexlab{b}.
\newblock \showarticletitle{{FlexTensor}: An Automatic Schedule Exploration and
  Optimization Framework for Tensor Computation on Heterogeneous System}. In
  \bibinfo{booktitle}{\emph{International Conference on Architectural Support
  for Programming Languages and Operating Systems (ASPLOS)}}.
  \bibinfo{pages}{859--873}.
\newblock
\urldef\tempurl%
\url{https://doi.org/10.1145/3373376.3378508}
\showDOI{\tempurl}


\bibitem[\protect\citeauthoryear{Zhou, Roy, Abdolrashidi, Wong, Ma, Xu, Liu,
  Phothilimtha, Wang, Goldie, Mirhoseini, and Laudon}{Zhou
  et~al\mbox{.}}{2020}]%
        {Zhou:2020:TGO}
\bibfield{author}{\bibinfo{person}{Yanqi Zhou}, \bibinfo{person}{Sudip Roy},
  \bibinfo{person}{Amirali Abdolrashidi}, \bibinfo{person}{Daniel Wong},
  \bibinfo{person}{Peter Ma}, \bibinfo{person}{Qiumin Xu},
  \bibinfo{person}{Hanxiao Liu}, \bibinfo{person}{Phitchaya Phothilimtha},
  \bibinfo{person}{Shen Wang}, \bibinfo{person}{Anna Goldie},
  \bibinfo{person}{Azalia Mirhoseini}, {and} \bibinfo{person}{James Laudon}.}
  \bibinfo{year}{2020}\natexlab{}.
\newblock \showarticletitle{Transferable Graph Optimizers for ML Compilers}. In
  \bibinfo{booktitle}{\emph{Advances in Neural Information Processing
  Systems}}, \bibfield{editor}{\bibinfo{person}{H.~Larochelle},
  \bibinfo{person}{M.~Ranzato}, \bibinfo{person}{R.~Hadsell},
  \bibinfo{person}{M.~F. Balcan}, {and} \bibinfo{person}{H.~Lin}} (Eds.),
  Vol.~\bibinfo{volume}{33}. \bibinfo{pages}{13844--13855}.
\newblock


\end{thebibliography}

\end{document}